\documentclass[]{jfm}
\usepackage{graphicx}
\usepackage{newtxtext}
\usepackage{newtxmath}
\usepackage{natbib}
\usepackage{hyperref}
\usepackage{bigints}
\usepackage{verbatim}
\hypersetup{
    colorlinks = true,
    urlcolor   = blue,
    citecolor  = black,
}

\newcommand{\RomanNumeralCaps}[1]
\linenumbers
\newcommand{\de}{\mathrm{d}}

\title{Heat transfer in drop-laden turbulence}
\author[F. Mangani, A. Roccon, F. Zonta, and A. Soldati]
{ 
Francesca Mangani$^{1}$,
Alessio Roccon$^{1,2}$,
Francesco Zonta$^{1}$ and
Alfredo Soldati$^{1,2}$
 \thanks{Email address for correspondence: alfredo.soldati@tuwien.ac.at}}

\affiliation{
$^1$ Institute of Fluid Mechanics and Heat Transfer, TU-Wien, 1060 Vienna, Austria
\\[\affilskip]
$^2$ Polytechnic Department of Engineering and Architecture, University of Udine, 33100 Udine, Italy}

\begin{document}
\graphicspath{{./figures/}}

\maketitle

\begin{abstract}

Heat transfer  by large deformable drops in a turbulent flow is a complex and rich in physics system, in which  drops deformation, breakage and coalescence influence the transport of heat. 
We study this problem coupling direct numerical simulations (DNS) of turbulence, with a phase-field method for the interface description. 
Simulations are run at fixed shear Reynolds and Weber numbers. 
To evaluate the influence of microscopic flow properties, like momentum/thermal diffusivity, on  macroscopic flow properties, like mean temperature or heat transfer rates, 
we consider four different values of the Prandtl number, which is  the  momentum to thermal diffusivity ratio: $Pr=1$, $Pr=2$, $Pr=4$ and  $Pr=8$.
The drops volume fraction is  $\Phi \simeq 5.4\%$ for all cases.
Drops are initially warmer than the turbulent carrier fluid, and release heat at different rates, depending on the value of $Pr$, but also on their size and on their own dynamics (topology, breakage, drop-drop interaction).
Computing the time behavior of the drops and carrier fluid average temperatures, we clearly show that an increase of $Pr$ slows down the heat transfer process. 
We explain our results by 
a simplified phenomenological model: we show  that the time  behavior of the drops average temperature is self similar, and a universal behavior can be found upon rescaling by $t/Pr^{2/3}$. Accordingly, the heat transfer coefficient $\mathcal{H}$ (resp. its dimensionless counterpart, the Nusselt number $Nu$) scales as $\mathcal{H}\sim Pr^{-2/3}$  (resp. $Nu\sim Pr^{1/3}$) at the beginning of the simulation,   and tends to $\mathcal{H}\sim Pr^{-1/2}$  (resp. $Nu\sim Pr^{1/2}$) at later times. These different scalings can be explained via the boundary layer theory and are consistent with previous theoretical/numerical predictions.

\end{abstract}



\section{Introduction}
\label{intro}

Transport of  passive and active scalars in multiphase turbulence is very important in many industrial processes and natural phenomena, from vaporization of atomized fuel jets \citep{gorokhovski2008modeling,ashgriz2011handbook,gao2022effect,boyd2023consistent}, to rain formation and atmosphere-ocean heat/mass exchanges \citep{duguid1971evaporation,deike2022mass} or even to the uptake of nutrients and other biochemicals by cells in complex flows \citep{aksnes1991theoretical,magar2005average}.
While the mixing of active or passive scalars in turbulent single-phase flows has been extensively analyzed  \citep{antonia2003effect,Kasagi1992,kim1989transport,warhaft2000passive,pirozzoli2016passive,Zonta2012,Zonta2012b,Zonta2014}, it remains a challenging task in  turbulent multiphase flows  \citep{gauding2022structure,rui2024annual}, where most of the available studies considered the case of heat/mass transfer from/to isolated drops and bubbles 
\citep{boussinesq1905calcul,levich1962,bird2002transport,bothe2004direct,figueroa2010mass}, with some remarkable exceptions -- focusing on  swarms of drops/bubbles -- appeared only recently \citep{herlina2016isotropic,albernaz2017droplet,dodd2021analysis,farsoiya2021bubble,shao2022interaction,scapin2022finite,hidman2023assessing,farsoiya2023direct}.
One of the crucial aspects  in turbulent multiphase flows, which makes these flows so difficult to analyze, is the presence of interfaces that dynamically move and deform in time and space according to the flow conditions, and that  mediate heat/species transport, mixing and phase change phenomena
\citep{deckwer1980mechanism,gvozdic2018experimental,liu2022heat,pelusi2023analysis}.

In this work, we focus  on the numerical simulation of  the heat transfer process in a droplet laden turbulent channel flow, looking in particular at the role of the Prandtl number $Pr$, i.e. the ratio between momentum and thermal diffusivity, in the process.
%
%
Compared to single-phase turbulence, where the range of scales that must be resolved to perform a direct numerical simulation (DNS) is purely dictated by the smallest scales of turbulence (Kolmogorov scale),  when the mixing of scalars in multiphase turbulence is analyzed, two further additional scales come into the picture.
The first one  is the Batchelor scale \citep{batchelor1959small1,batchelor1959small2}, which determines the smallest scale of the temperature/concentration field.
The second important scale  is the Kolmogorov-Hinze scale \citep{kolmogorov1941local,Hinze1955}, and is  linked to the multiphase nature of the flow.
This scale can be used, perhaps with some limitations \citep{qi2022fragmentation}, to determine the critical size of a drop/bubble that will not undergo breakage in turbulence.
These two scales -- and their corresponding ratio to the Kolmogorov scale, i.e. the smallest length scale of the turbulent flow field -- control the system dynamics and define the minimal grid requirements that must be satisfied to perform a DNS of scalar mixing in multiphase turbulence (keeping always in mind that performing a simulation that resolves the interface dynamics down to the molecular scale is at present almost unfeasible).
In this context, the major constraint is usually posed by the Batchelor scale, which becomes smaller than the Kolmogorov length scale when Prandtl numbers  large than unity are considered.  
Overall, the wide range of scales involved in the process makes simulations  of scalar mixing in multiphase turbulence a challenging task and limits the space parameters that can be explored by means of direct numerical simulations.
Our simulations are initialized injecting a swarm of large and deformable drops (initially warmer) inside a turbulent channel flow (initially colder).
The  system is described by coupling the direct numerical simulations of turbulent heat transfer with a phase-field method, employed to describe the drops topology  \citep{zheng2015phase,mirjalili2022computational}.
We simulate realistic values of the Prandtl number up to $Pr=8$, similar to those obtained in liquid-liquid systems.
We remark here that simulations of mass transfer problems in wall-bounded flow configurations, where the typical Schmidt number $Sc$ (i.e. the mass transfer counterpart of $Pr$) is $\mathcal{O}({10^2 \sim 10^3})$, e.g. $Sc \simeq 600$ for $CO_2$ in freshwater \citep{wanninkhof1992relationship}, are currently out of reach even using the most advanced computing.
Indeed, the resulting Batchelor scale would be at least one order of magnitude smaller,  thus requiring grid resolutions comparable or larger than those employed for state-of-the-art single-phase DNS \citep{leemoser2015,pirozzoli2021one} but with a much larger computational cost as the systems of equations to be solved is more complex and restrictive (also from the temporal discretization point of view).

The present study has three main objectives. 
First, we want to investigate the macroscopic dynamic of the drops and of the heat transfer process by analyzing the drop size distribution and the mean temperature behavior of the two phases over time.
Second, we want to characterize the influence  of the Prandtl number, i.e. of the microscopic flow properties, on the macroscopic flow properties (mean temperature, heat transfer coefficient) and, building on top of the numerical results, we want to develop a physically-based model  to explain the observed results.
Third, we want to study the influence of the Prandtl number and of drop size on the temperature distribution inside the drops, so to evaluate the corresponding flow mixing/ homogenization.

The paper is organized as follows. 
In section \S2, the governing equations, the numerical method, and the simulation setup are presented.
In section \S3, the simulation results, in terms of drop size distribution and mean temperature of the two phases and heat transfer coefficient are carefully characterized and discussed. 
A simplified model is also developed to explain the observed results.  
%
The temperature distribution inside the drops is then evaluate at  different Prandtl numbers and drop sizes.
Finally, conclusions are presented in §4.

\section{Methodology}
\label{met}

We consider a swarm of large and deformable drops injected in a turbulent channel flow.
The channel has dimensions $L_x \times L_y \times L_z = 4 \pi h \times 2 \pi h \times 2h$ along the streamwise ($x$), spanwise ($y$) and wall-normal direction ($z$).
To describe the dynamics of the system, we couple direct numerical simulation (DNS) of the Navier-Stokes and energy equations, used to describe the turbulent flow, with a phase-field method (PFM), used to describe the interfacial phenomena.
%
%
The employed numerical framework is described more in detail in the following.
%

\subsection{Phase-field method}
To describe the dynamics of drops and the corresponding   topological changes (e.g. coalescence and breakage), we employ an energy-based phase field method \citep{Jacqmin1999,Badalassi2003}, which is based on the introduction of a scalar quantity, the phase field $\phi$, required  to identify the two phases. 
%
%
The phase field $\phi$ has a uniform value in the bulk of each phase ($\phi=+1$ inside the drops; $\phi=-1$ inside the carrier fluid) and undergoes a smooth change across the thin transition layer that separates the two phases.
The transport of the phase field variable is described by a Cahn-Hilliard equation, which in dimensionless form reads as:
\begin{equation}
\frac{\partial \phi}{\partial t} + {\bf u} \cdot \nabla \phi  = \frac{1}{\Pen} \nabla^2 \mu_\phi \, + f_p ,
\label{ch}
\end{equation}
where ${\bf u}=(u,v,w)$ is the velocity vector, $\Pen$ is the P\'eclet number, $\mu$ is the phase field chemical potential while $f_p$ is a penalty-flux term which will be further discussed later.
%
The P\'eclet number is
\begin{equation}
\Pen=\frac{u_\tau^* h^* }{{\mathcal{M}^*} \beta^*} \, ,
\end{equation}
where $u^*_\tau$ is the friction velocity ($u_\tau^*=\sqrt{\tau_w^*/\rho^*}$, with  $\tau_w^*$ the wall-shear stress and $\rho^*=\rho_c^*=\rho_d^*$ the density of the fluids), $h^*$ is the channel half-height, ${\mathcal{M}}^*$ is the mobility and $\beta^*$ is a positive constant (the superscript $^*$ is used to denote dimensional quantities hereinafter).
%
%
%
The chemical potential $\mu$ is defined as the variational derivative of a Ginzburg-Landau free-energy functional, the expression of which is chosen to represent an immiscible binary mixture of fluids 
%
%
\citep{Soligo2019c,Soligo2019a,Soligo2019b}.
The functional is the sum of two contributions: the first contribution, $f_0$, accounts for the tendency of the system to separate into the two pure stable phases, while the second contribution, $f_{mix}$, is a mixing term accounting for the energy stored at the interface (i.e. surface tension). 
The mathematical expression of the functional in dimensionless form is:
\begin{equation}
\mathcal{F}[\phi, \nabla \phi]=\bigintsss_{\Omega} \bigg( \underbrace{\frac{({\phi}^2-1)^2}{4}}_{f_0}+\underbrace{\frac{Ch^2}{2} \left| \nabla \phi \right|^2}_{f_{mix}} \bigg)  \de \Omega \, ,
\label{ginz-land}
\end{equation}
where $\Omega$ is the considered domain  and $Ch$ is the Cahn number, which represents  the dimensionless thickness of the thin interfacial layer between the two fluids:
\begin{equation}
Ch=\frac{\xi^*}{h^*} \, ,
\end{equation} 
where $\xi^*$ is clearly the dimensional thickness of the interfacial layer.
From equation~(\ref{ginz-land}), the expression of the chemical potential can be derived as the functional derivative with respect to the order parameter:
\begin{equation}
\mu_\phi=\frac{\delta \mathcal{F}[\phi \nabla \phi ]}{\delta \phi}={\phi}^3 - \phi - Ch^2  \nabla^2 \phi \, .
\end{equation}
At equilibrium, the chemical potential is constant throughout all the domain. 
The equilibrium profile for a flat interface can thus be obtained by solving $\nabla \mu_\phi = \mathbf{0}$, hence:
\begin{equation}
\phi_{eq}=\tanh \left( \frac{s}{\sqrt{2}Ch}\right)\, ,
\end{equation}
where $s$ is the coordinate normal to the interface.
%
As anticipated before, the last term in the right-hand side of the Cahn-Hilliard equation~(equation \ref{ch}) is a penalty-flux term employed in the profile-corrected formulation of the phase-field method, and is used to overcome some potential drawbacks of the standard formulation of the method, e.g. mass leakages among the phases and misrepresentation of the interfacial profile \citep{YUE2007,li2016phase}.
%
This penalty flux is defined as:
\begin{equation}
f_p=\frac{\lambda}{Pe} \left[ \nabla^2 \phi -\frac{1}{\sqrt{2}Ch} \nabla \cdot  \left( (1-{\phi}^2) \frac{\nabla\phi}{|\nabla\phi|} \right) \right] \, ,
\end{equation}
where  $\lambda=0.0625/Ch$  \citep{Soligo2019b}.

\subsection{Hydrodynamics}
\label{hydro}

To describe the hydrodynamics of the multiphase system, the Cahn-Hilliard equation is coupled with the Navier-Stokes equations.
The presence of a deformable interface (and of the corresponding surface tension forces) is accounted for by introducing an interfacial term in the Navier-Stokes equations. 
Recalling that in the present case we consider two fluids with the same  density ($\rho^*=\rho_c^*=\rho_d^*$) and viscosity ($\mu^*=\mu_c^* =\mu_d^*$), continuity and Navier-Stokes equations in dimensionless form read as:
\begin{equation}
\nabla \cdot \mathbf{u}=0 \, ,
\label{cont}
\end{equation}
\begin{equation}
  \frac{\partial \mathbf{u}}{\partial t}+\mathbf{u} \cdot \nabla \mathbf{u} =
-\nabla p +\frac{1}{Re_\tau} \nabla^2 \mathbf{u} +  \frac{3}{\sqrt{8}}\frac{Ch}{We} \nabla \cdot \mathbf{T_c} \, ,
\label{ns}
\end{equation}
%
%
$p$ is the pressure field, while  $\mathbf{T_c}$ is the Korteweg tensor \citep{KORTEWEG1901} used to account for the surface tension forces and defined
\begin{equation}
\mathbf{T_c}=|\nabla\phi |^2 \mathbf{I}-\nabla\phi \otimes \nabla \phi \, ,
\end{equation}
where $\mathbf I$ is the identity matrix and $\otimes$ represents the dyadic product.
This approach is the continuum surface stress approach \citep{lafaurie1994modelling,gueyffier1999volume} applied in the context of PFM, and is analytically equivalent to the chemical potential forcing \citep{mirjalili2023assessment}.
%
The dimensionless groups appearing in the Navier-Stokes equations are the shear Reynolds number $Re_\tau$ (ratio between inertial and viscous forces), and the Weber number $We$ (ratio between inertial and surface tension forces), which are defined as:
\begin{equation}
Re_\tau=\frac{\rho^* u_\tau^* h^*}{\mu^*}\, ,  \qquad
We=\frac{\rho^* {{u_\tau^*}}^2  h^*}{\sigma^*}\,.
\label{dp}
\end{equation}
%
%
where $\sigma^*$ is the surface tension. 
Note that, consistently with the employed adimensionalization,   $We$ is defined  using the half-channel height (and not the drop diameter). 
%

\subsection{Energy equation}
\label{thermal}

The time evolution of the temperature field is obtained by solving the energy equation using a one-scalar model approach \citep{zheng2015phase}.
%
%
To avoid the introduction of further complexity in the system, we consider two fluids with the same thermophysical properties, i.e. same thermal conductivity $\lambda^*$, same specific heat capacity $c_p^*$ and therefore same thermal diffusivity $a^*=\lambda^*/\rho^* c_p^*$ (since the density of the two phases is also the same).
These properties have been evaluated at a reference temperature $\theta_{r}^*=(\theta_{d,0}^* + \theta_{c,0}^*)/2$, i.e. the average between the initial drops temperature and the carrier fluid temperature,  and are assumed to be constant and uniform.
Within these assumptions, the energy equation written in dimensionless form reads as:
\begin{equation}
\frac{\partial \theta }{\partial t} + {\bf u} \cdot \nabla \theta  = \frac{1}{\Rey_\tau Pr} \nabla^2 \theta \, ,
\label{temp}
\end{equation}
%
where $Pr$ is the Prandtl number defined as
\begin{equation}
Pr=\frac{\mu^* c_p^*}{\lambda^*}=\frac{\nu^*}{a^*}\, ,
\end{equation}
with $\nu^*=\mu^*/\rho^*$  the  kinematic viscosity (i.e. momentum diffusivity). 
%
From a physical viewpoint, 
$Pr$ represents the  momentum-to-thermal diffusivity ratio.
\subsection{Numerical discretization}

The governing equations~(\ref{ch}), (\ref{cont}), (\ref{ns}) and~(\ref{temp}) are solved using a pseudo-spectral method, which uses Fourier series along the periodic directions (streamwise and spanwise) and Chebyshev polynomials along the wall-normal direction.
The Navier-Stokes and continuity equations are solved using a wall-normal velocity-vorticity formulation: equation~(\ref{ns}) is rewritten as a $4^{th}$ order equation for the wall-normal component of the velocity $w$ and a $2^{nd}$ order equation for the wall-normal component of the vorticity $\omega_z$ \citep{Kim1987,Speziale1987}. 
The Cahn-Hilliard equation~(\ref{ch}), which in its original form is a $4^{th}$ order equation is split into two $2^{nd}$ order equations using the splitting scheme proposed in \cite{Badalassi2003}.
Using this scheme, the governing equations are recasted as a coupled system of Helmholtz equations, which can be readily solved.

The governing equations are time advanced using an implicit-explicit scheme. 
For the Navier-Stokes, the linear part is integrated using a Crank-Nicolson implicit scheme, while the non-linear part is integrated using an Adams-Bashforth explicit scheme.
Similarly, for the Cahn-Hilliard and energy equations, the linear terms are integrated using an implicit Euler scheme, while the non-linear terms are integrated in time using an Adams-Bashforth scheme. 
The adoption of the implicit Euler scheme for the Cahn-Hilliard equation helps to damp unphysical high-frequency oscillations that could arise from the steep gradients of the phase field. 

As the characteristic length scales of the flow and temperature fields, represented by the Kolmogorov scale, $\eta_k^+$, and the Batchelor scale, $\eta_\theta^+$, are different when non-unitary Prandtl numbers are employed (being these two quantities linked by the following relation $\eta_\theta^+=\eta_k^+/\sqrt{Pr}$), a dual grid approach is employed to reduce the computational cost of the simulations and at the same time to fulfill the DNS requirements.
In particular, when super-unitary Prandtl numbers are simulated, a finer grid is used to resolve the energy equation.
Spectral interpolation is used to upscale/downscale the fields from the coarse to the refined grid and vice-versa when required (e.g. upscaling of the velocity field to compute the advection terms in the energy equation).

This numerical scheme has been implemented in a parallel Fortran 2003 MPI in-house proprietary code.
The parallelization strategy is based on a 2D domain decomposition to divide the workload among all the MPI tasks.
The solver execution is accelerated using openACC directives and CUDA Fortran instructions (solver execution) while the Nvidia cuFFT libraries are used to accelerate the execution of the Fourier/Chebyshev transforms. 
Overall, the computational method adopted allows for the accurate resolution of all the governing equations and the achievement of an excellent parallel efficiency thanks to the fine-grain parallelism offered by the numerical method used.
The equivalent computational cost of the simulations is about 25 million CPU hours and the resulting dataset has a size of about 16 TB.

\subsection{Boundary conditions}

The system of governing equations is complemented by a set of suitable boundary conditions.
For the Navier-Stokes equations, no-slip boundary conditions are enforced at the top and bottom walls (located at $z=\pm h$):
\begin{equation}
{\bf u} (z=\pm h)=\mathbf{0} \, .
\end{equation}
For the Cahn-Hilliard equation,  no-flux boundary conditions are applied at the two walls, yielding to the following boundary conditions:
\begin{equation}
\frac{\partial \phi}{\partial z}(z=\pm h)=0 \, ; \qquad
\frac{\partial^3 \phi}{\partial {z}^3}(z=\pm h)=0 \, .
\end{equation}
Likewise, for the energy equation, no-flux boundary conditions are applied at the two walls (i.e. adiabatic walls).
\begin{equation}
\frac{\partial \theta}{\partial z}(z=\pm h)=0 \, .
\end{equation}

Along the streamwise and spanwise directions ($x$ and $y$), periodic boundary conditions are imposed for all variables (Fourier discretization).
The adoption of these boundary conditions leads to the conservation of the phase field and temperature fields over time:
\begin{equation}
\frac{\partial}{\partial t} \int_{\Omega}^{} \phi \de \Omega = 0\, ; \qquad
\frac{\partial}{\partial t} \int_{\Omega}^{}\theta \de \Omega = 0 \, .
\label{cpfm}
\end{equation}
where $\Omega$ is the computational domain.
Regarding the phase-field, equation~(\ref{cpfm}) enforces mass conservation of the entire system but does not guarantee the conservation of the mass of each phase \citep{YUE2007,Soligo2019b}, as some leakages between the phases may occur. 
This drawback is rooted in the phase-field method and more specifically in the curvature-driven flux produced by the chemical potential gradients \citep{Kwakkel2020,Mirjalili2021}.
This issue is here successfully mitigated with the adoption of the profile-corrected formulation that largely reduces this phenomenon.
In the present cases, mass leakage between the phases occurs only in the initial transient when the phase field is initialized (see the section below for details on the initial condition) and is limited to 2\% of the  initial mass of the drops.
After this initial transient, the mass of each phase keeps constant. 


\subsection{Simulation set-up}
\label{sec: setup}

The turbulent channel flow, driven by an imposed constant pressure gradient in the streamwise direction, has shear Reynolds number $Re_\tau=300$.
The computational domain has dimensions $L_x \times L_y \times L_z= 4\pi h   \times 2\pi h \times 2h$, which corresponds to $L_x^+ \times L_y^+ \times L_z^+=3770 \times 1885 \times 600$ wall units.
%
The value of the  Weber number is kept constant and is equal to $We=3.00$, so to be  representative of liquid/liquid mixtures \citep{Than1988}. 
To study the influence of the Prandtl number $Pr$ on the heat transfer process, we consider four different values of $Pr$: $Pr=1$, $Pr=2$, $Pr=4$ and $Pr=8$.
These values cover a wide range of real-case scenarios: from low-Prandtl number fluids to water-toluene mixtures.

The grid resolution used to resolve the continuity, Navier-Stokes and Cahn-Hilliard equations is equal to $N_x \times N_y \times N_z = 1024 \times 512 \times 513$ for all the cases considered in this work.
For the energy equation, the same grid used for the flow field and phase field is employed at the lower Prandtl numbers ($Pr=1$ and $Pr=2$), while a more refined grid, with $N_x \times N_y \times N_z = 2048 \times 1024 \times 513$ points, is used when the larger Prandtl numbers are considered ($Pr=4$ and $Pr=8$). 
The computational grid has uniform spacing in the homogeneous directions, while Chebyshev-Gauss-Lobatto points are used in the wall-normal direction. 
We refer the reader to table~\ref{tab1} for an overview of the main phzsical and computational parameters of the simulation.
For the employed grid resolution,  the Cahn number is set to $Ch = 0.01$ while, to achieve convergence to the sharp interface limit, the corresponding phase field P\'eclet number is $\Pen = 1/Ch = 50$.
%
%
%

\begin{table}
\centering
\begin{tabular}{c ccccc}
Case& $\Rey_\tau$  & $We$ & $Pr$ &  $N_x \times N_y \times N_z$ (NS+CH) &  $N_x \times N_y \times N_z$ (Energy) \\
\hline
Single-phase    &300& -&-& $512 \times 256 \times 257$ & - \\
Drop-laden &300&3.0&1.0&$1024 \times 512 \times 513$& $1024 \times 512 \times 513$ \\
Drop-laden &300&3.0&2.0& $1024 \times 512 \times 513$& $1024 \times 512 \times 513$ \\
Drop-laden &300&3.0&4.0& $1024 \times 512 \times 513$& $2048 \times 1024 \times 1025$ \\
Drop-laden &300&3.0&8.0&$1024 \times 512 \times 513$& $2048\times 1024 \times 1025$ \\
\end{tabular}
\caption{Overview of the simulation parameters.
For a fixed shear Reynolds number $Re_\tau=300$ and Weber number $We=3$, we consider a single-phase flow case and four non-isothermal drop-laden flows characterised by different Prandtl numbers: from $Pr=1$ to $Pr=8$. 
The grid resolution is modified accordingly so as to satisfy DNS requirements.}
\label{tab1}
\end{table}

All simulations are initialized releasing a regular array of 256  spherical drops with diameter $d = 0.4h$ (corresponding to $d^+ = 120~w.u.$) inside a fully-developed turbulent flow field (obtained from a preliminary simulation).
To ensure the independence of the results from the initial flow field condition, each case is initialized with a slightly different flow field realization.
Naturally, the fields are equivalent in terms of statistics as they are all obtained from a statistically steady turbulent channel flow.
The volume fraction of the drops is $\Phi = V_d/(V_c + V_d) = 5.4 \%$, with $V_d$ and $V_c$ the volume of the drops and carrier fluid, respectively.

\begin{figure}
\begin{picture}(300,270)
\put(0,0){\includegraphics[width=1.0\columnwidth]{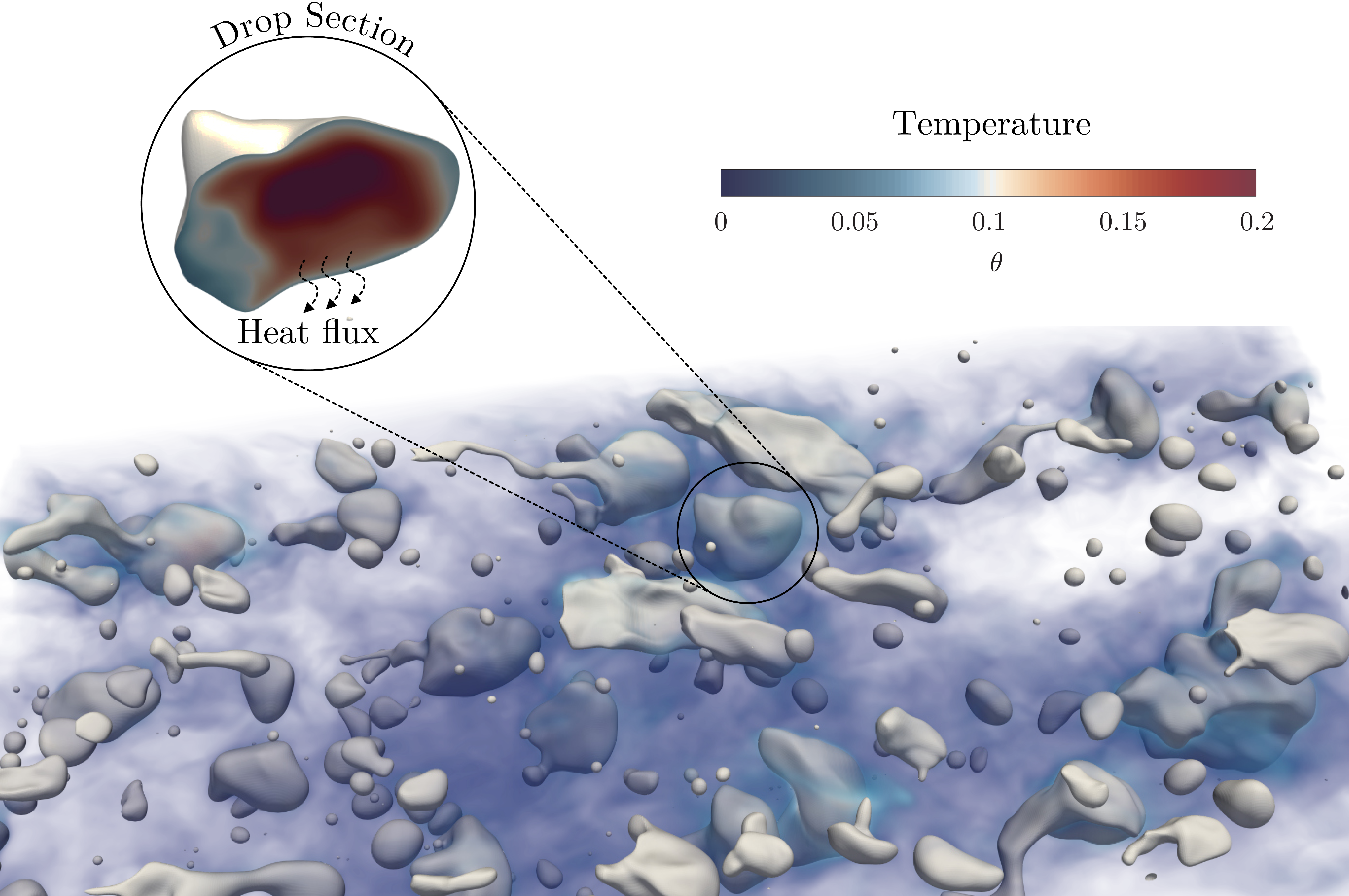}}
\end{picture}
\caption{Rendering of the computational setup employed for the simulations.
A swarm of large and deformable drops is released in a turbulent channel flow.
The flow goes from left to right driven by a constant pressure gradient.
The temperature field is volume-rendered (blue-low, red-high) and the drops interface is shown in white (iso-level $\phi=0$).
As can be appreciated from the close-up view (top left, which shows the temperature field in a drop section), drops have a temperature higher than the carrier fluid and as a result, there is a heat flux from the drops to carrier fluid.
The snapshot refers to $Pr=1$ and $t^+=1000$. }
\label{introp}
\end{figure}

The initial condition for the temperature field is such that all drops are initially warm (initial  temperature $\theta_{d,0}=1$), while the carrier fluid is initially cold (initial temperature $\theta_{c,0}=0$).
To avoid numerical instabilities that might arise from a discontinuous temperature field, the transition between drops and carrier fluid is initially smoothed  using a hyperbolic tangent kernel.
Figure~\ref{introp} (which is an instantaneous snapshot captured at $t^+=1000$, for $Pr=1$) shows a volume rendering of the  temperature field (blue-cold, red-hot), inside which deformable drops (whose interface, iso-level $\phi=0$, is shown in white) are transported.

\section{Results}

Results obtained from the numerical simulations will be first discussed from a qualitative viewpoint, by looking at instantaneous flow and drops visualizations, and then analyzed from a more quantitative viewpoint, by looking at the drop size distribution (DSD), and at the effect of the Prandtl number $Pr$ on the average drops and fluid temperature. 
To explain the numerical results, and to offer a possible parametrization of the heat transfer process in drop-laden flows, we will also develop a simplified phenomenological model of the system.
%
%
Finally, we will characterize the temperature distribution inside the drops, elucidating the effects of $Pr$ and of the drop size on it.
Note that, unless differently mentioned, results are presented using the wall-unit scaling system but for the temperature field, which is made  dimensionless using the initial temperature difference as a reference scale (which is a natural choice in the present case).

\subsection{Qualitative discussion}

\begin{figure}
\begin{picture}(400,220)
\put(170,202){\bf Breakage}
\put(12,190){$(a)~t^+=600$}
\put(90,190){$(b)~t^+=645$}
\put(165,190){$(c)~t^+=660$}
\put(243,190){$(d)~t^+=675$}
\put(320,190){$(e)~t^+=690$}
\put(168,106){\bf Coalescence}
\put(12,94){$(f)~t^+=1515$}
\put(90,94){$(g)~t^+=1530$}
\put(165,94){$(h)~t^+=1545$}
\put(243,94){$(j)~t^+=1560$}
\put(320,94){$(k)~t^+=1575$}
\put(183,8){Time}
\put(0,-10){\includegraphics[width=1.0\columnwidth]{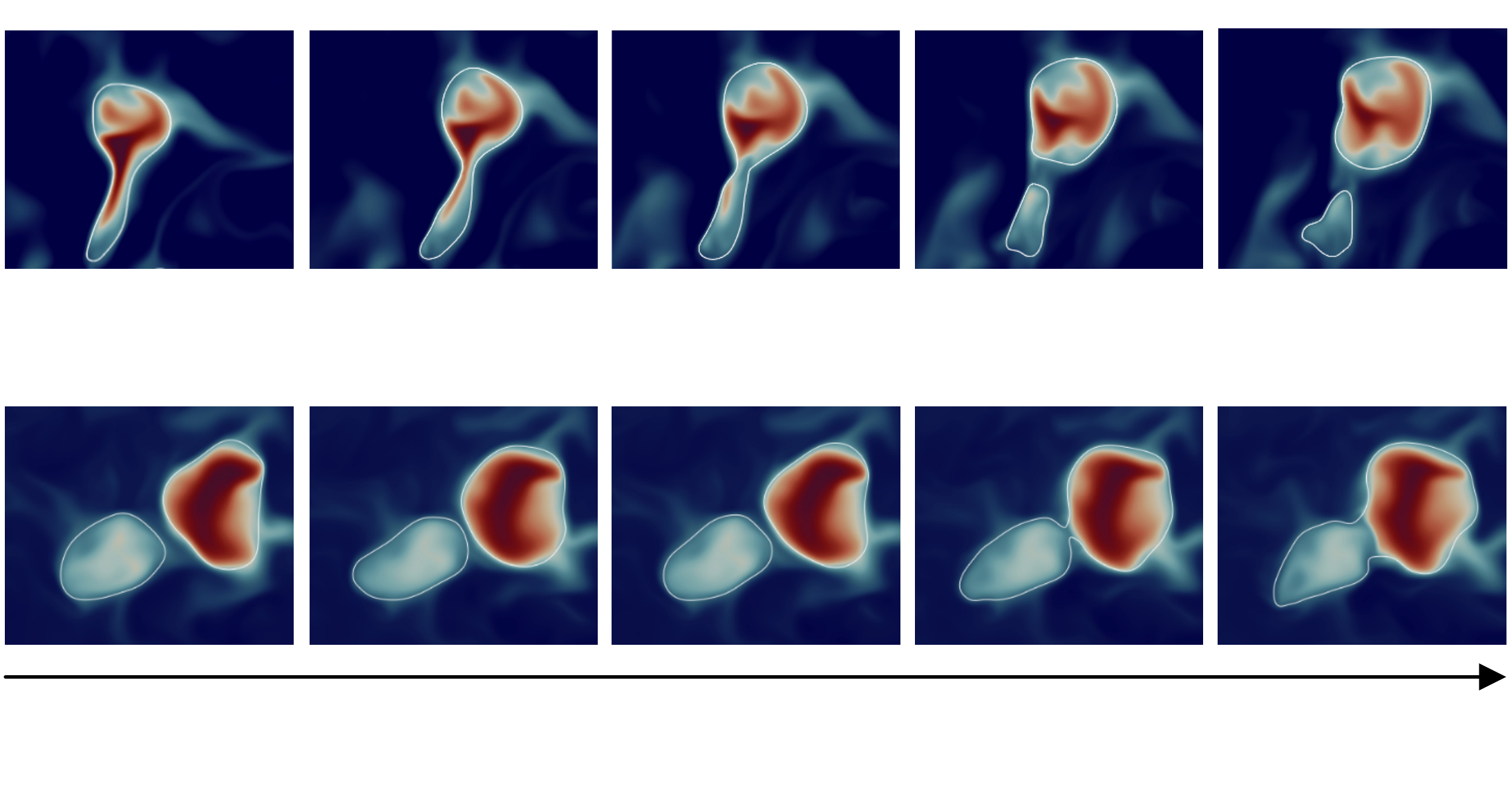}}
\end{picture}
\caption{Influence of topology changes on heat transfer: Time sequence of a breakage event (top row, panels~$a$-$e$) and of a coalescence event  (bottom row, panels~$f$-$k$).
In the top row, we can appreciate how during a breakage event in the pinch-off region heat is efficiently transferred from the drops to carrier fluid thanks to the high surface/volume ratio of this region. 
Likewise, the thermal relaxation time of the small drops generated during the breakage (bottom) is smaller than that of the parent drop (top).
In the bottom row, we can appreciate how during a coalescence event, parcels of fluid with different temperatures mix.
The two time sequences refer to the case $Pr=1$ and snapshots are separated by $\Delta t^+ =15$. }
\label{time}
\end{figure}
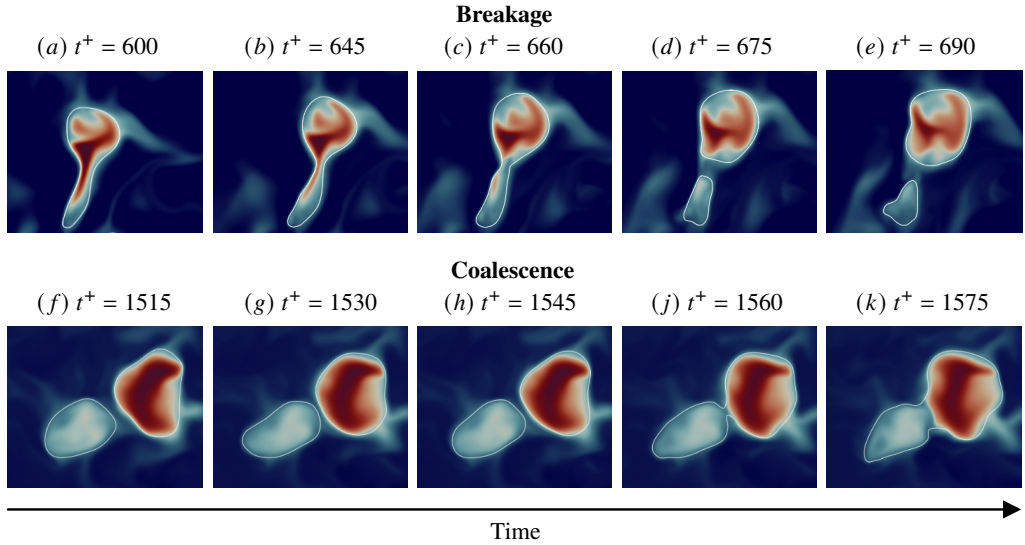

The complex dynamics of drops immersed in a non-isothermal turbulent flow is  visualized in figure~\ref{introp}, where the drops (identified  by isocontour of $\phi=0$) are shown together with a volume-rendered distribution of temperature in the carrier fluid. Also shown in figure~\ref{intro} is a close-up view of the temperature distribution inside the drop.

%
%
Once injected into the flow, each   drop starts interacting with the flow and with the neighbouring drops.  
The result of the drop-turbulence and drop-drop interactions is the occurrence of breakage and coalescence events.
A breakage event happens when the flow vigorously stretches the drop, leading to the formation of a thin ligament that breaks and generates two child drops. Upon separation, surface tension forces tend to retract the broken filaments and to restore the drop spherical shape.
A coalescence event is observed when two drops come close each other. The small liquid film that separates the drops start to drain, and a coalescence bridge is formed.
Later, surface tension forces enter the picture,  reshaping the drop and completing the coalescence process.
%
%
%
%
%
%
The dynamic competition between breakage and coalescence events, and their interaction with the turbulent flow, determines the number of drops,  their size distribution, and  their shape/morphology  (i.e., curvature, interfacial area, etc.). 
%

\begin{figure}
\setlength{\unitlength}{0.0025\columnwidth}
\begin{picture}(300,500)
\put(0,460){$Pr=1$}
\put(0,335){$Pr=2$}
\put(0,210){$Pr=4$}
\put(0,85){$Pr=8$}
\put(40,505){$(a)$}
\put(40,380){$(b)$}
\put(40,255){$(c)$}
\put(40,130){$(d)$}
\put(40,360){\includegraphics[width=0.7\columnwidth, keepaspectratio]{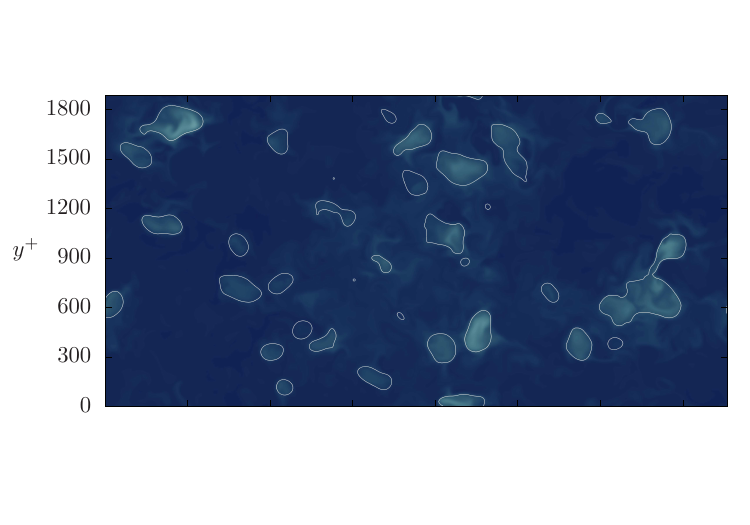}}
\put(40,236){\includegraphics[width=0.7\columnwidth, keepaspectratio]{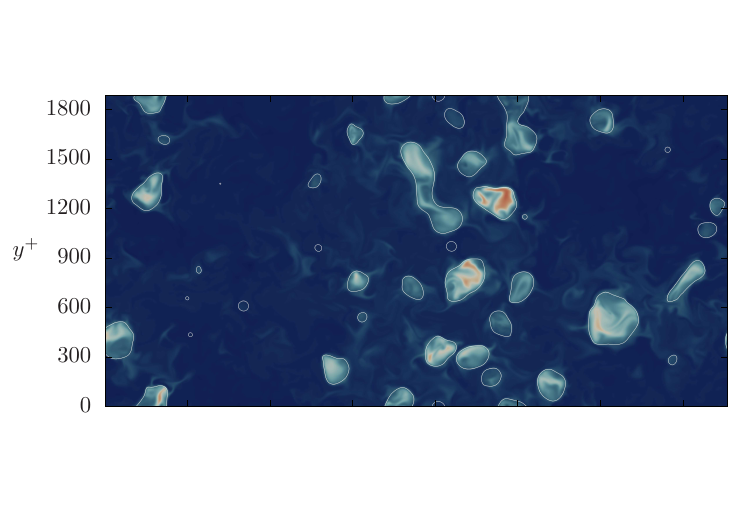}}
\put(40,111){\includegraphics[width=0.7\columnwidth, keepaspectratio]{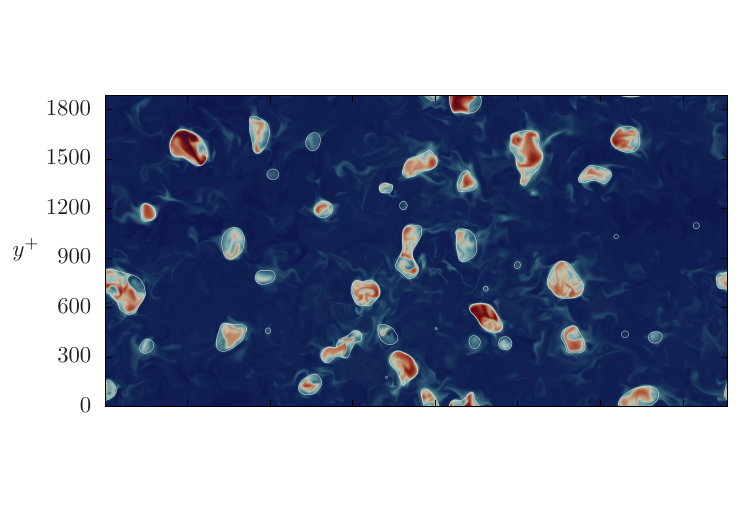}}
\put(40,-20){\includegraphics[width=0.7\columnwidth, keepaspectratio]{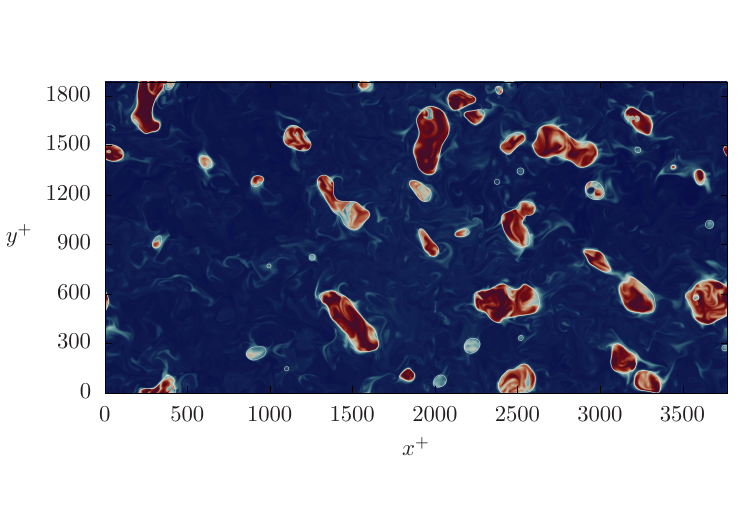}}
\put(210,180){\includegraphics[width=0.7\columnwidth, keepaspectratio]{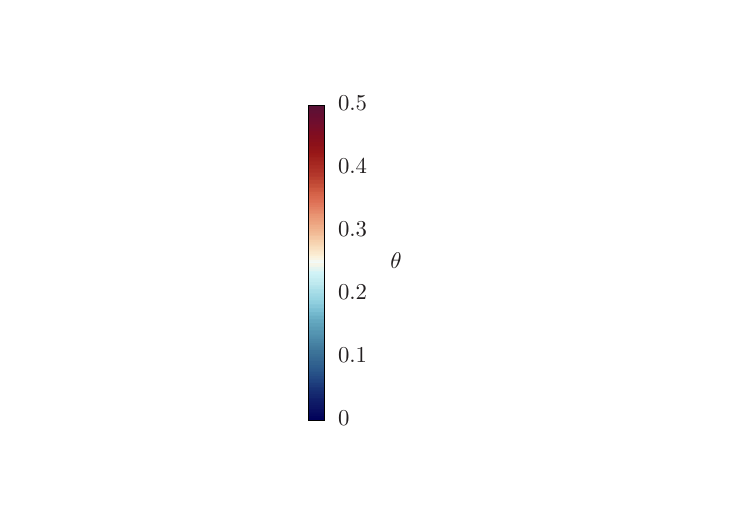}}
\end{picture}
\caption{Instantaneous visualization of the temperature field (red-hot; blue-cold) on a $x^+-y^+$ plane located at the channel center for $t^+=1500$. 
Drops interface (iso-level $\phi=0$) are reported using white lines.
Each panel refers to a different Prandtl number: $Pr=1$ (panel $a$), $Pr=2$ (panel $b$), $Pr=4$ (panel $c$) and $Pr=8$ (panel $d$).
By increasing the Prandtl number (from top to bottom) and thus decreasing the thermal diffusivity, the heat transfer process slows down.
As a consequence, the drop temperature is higher when larger Prandtl numbers are considered.
The effects of the Prandtl number on the characteristic length scales can also be appreciated: as it increases, temperature structures become smaller.}
\label{quali}
\end{figure}
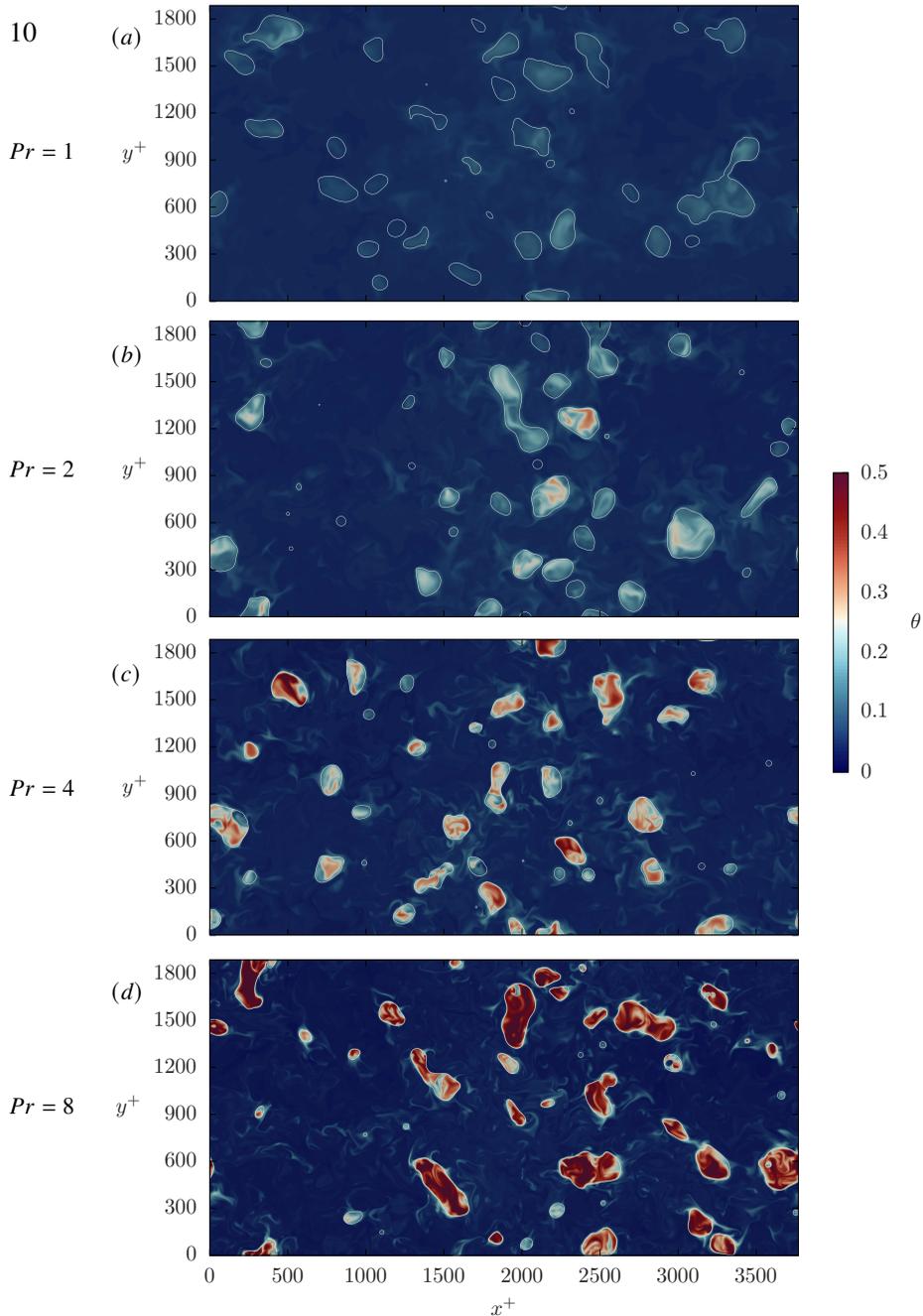

In the present case, drops not only exchange momentum with the flow and with the other drops, but also heat.
%
%
Starting from an initial condition  characterized by warm drops (with uniform temperature) and cold carrier fluid, and because of the imposed adiabatic boundary conditions, the system evolves towards an equilibrium isothermal state.
During the transient to attain this thermal equilibrium state,  heat is transported by diffusion  and advection  inside each of the two phases, and across the interface of the drops (see the temperature field inside and outside the drops, figure~\ref{intro}). 
The picture is then further complicated by the occurrence of  breakage and coalescence events.  This is represented  in figure \ref{time}. 
When a breakage occurs (figure \ref{time}, top row), a thin filament is generated (figure \ref{time}a-c), which then leads to the formation of a smaller  satellite drop (figure \ref{time}d-e). The filament and the satellite drop, given the large surface-to-volume ratio, exchange heat very efficiently, and become  rapidly colder.
In contrast, when a coalescence occurs (figure \ref{time}, bottom row), two drops having different temperature merge together. This induces an efficient mixing process, during which cold parcels of one drop become warmer and viceversa, warm parcles of the other drops become colder.
Overall, breakup and coalescence events induce heat transfer modifications that are in general  hard to predict a priori, since they do depend on the relative size of the involved parents/child  drops. 

Naturally, the problem of heat transfer in drop-laden turbulence is strongly influenced by the Prandtl number of the flow.
This can be appreciated by looking at figure~\ref{quali}, where we show the instantaneous temperature field, together with the shape of the drops, at a certain instant in time ($t^+=1500$) and  at the different Prandtl numbers: $Pr=1$ (panel $a$), $Pr=2$ (panel $b$), $Pr=4$ (panel $c$) and $Pr=8$ (panel $d$).
%
%
In each panel, the temperature field is shown on a wall-parallel $x^+$-$y^+$ plane located at the channel center ($z^+=0$), and is visualized with a blue-red scale (blue-low, red-high).
We observe that the temperature field changes significantly with  $Pr$. 
In particular, we notice an increase in the drop-to-fluid temperature difference for increasing $Pr$, going from $Pr=1$ (top panel) where this difference is small, to $Pr=8$ (bottom panel) where this difference is large.
The heat transfer from the drops to the carrier fluid becomes slower as $Pr$ increases, consistently with a  physical situation in which the $Pr$ number is increased by reducing the thermal diffusivity of the fluid, while keeping the momentum diffusivity constant (i.e. constant kinematic viscosity, and hence shear Reynolds number). 
%
%
%
Also, the temperature structures, both inside and outside the drop,  become thinner and more complicated at higher $Pr$, since their characteristic lengthscale, the Bachelor scale $\eta_\theta^+ \propto {Pr}^{-1/2}$, becomes smaller for increasing $Pr$ \citep{batchelor1959small1,batchelor1959small2}.
In addition, smaller drops have, on average, a lower temperature compared to larger drops, regardless of the value of $Pr$.
All these aspects will be discussed in more detail in the next sections.


\subsection{Drop Size Distribution} 

To characterize the collective dynamics of the drops, we compute the drop size distribution (DSD) at steady-state conditions, averaging over a time window $\Delta t^+=3000$, from $t^+=3000$ to $6000$.
It is worth mentioning that a quasi-equilibrium DSD, very close to the steady one, is already achieved around $t^+ \simeq 750$, and only minor changes occur to the DSD afterwards.

\begin{figure}
\begin{picture}(300,200)
\put(0,0){\includegraphics[width=0.9\columnwidth]{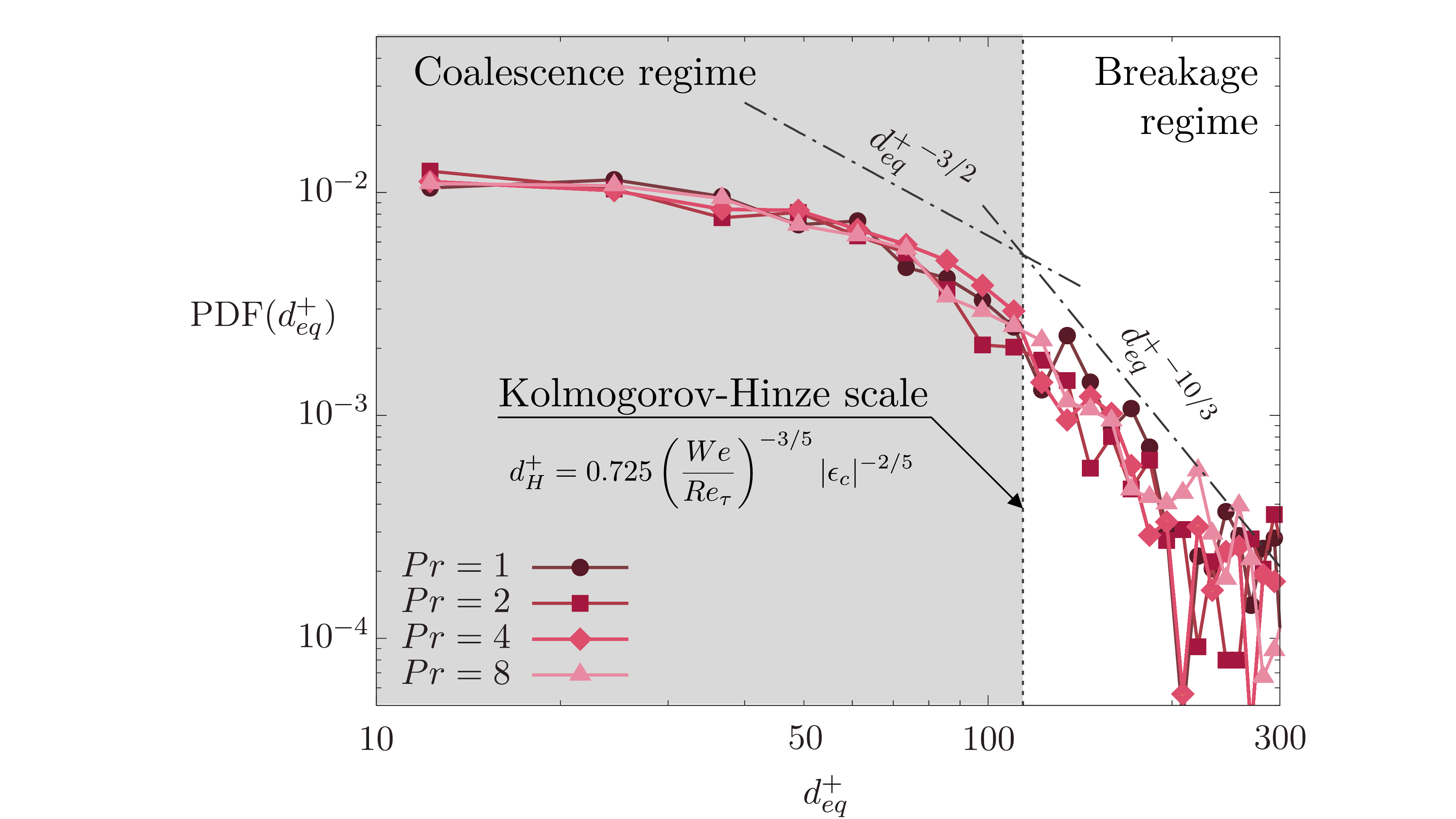}}
\end{picture}
\caption{Steady-state drop size distributions (DSD) obtained for: $Pr=1$ (dark violet, circles), $Pr=2$ (violet, squares), $Pr=4$ (pink, diamonds) and $Pr=8$ (light pink, triangles).
The Kolmogorov-Hinze (KH) scale $d^+_H$ is reported with a vertical dashed line while the two analytical scaling laws: ${d_{eq}^+}^{-3/2}$  for the coalescence-dominated regime (small drops, gray region) and ${d_{eq}^+}^{-10/3}$ for the breakage-dominated regime (larger drops, white region) are reported with dash-dotted lines.}
\label{dsd}
\end{figure}

Figure~\ref{dsd} shows the DSD obtained for the different cases considered here: $Pr=1$ (dark violet), $Pr=2$ (violet), $Pr=4$ (pink), and $Pr=8$ (light pink).
The distributions have been computed considering, for each drop, the diameter of the equivalent sphere computed as:
\begin{equation}
d_{eq}^+= \left( \frac{6 V^+}{\pi} \right)^{1/3}\, ,
\end{equation}
where $V^+$ is the volume of the drop.
Also reported in figure~\ref{dsd} is the Kolmogorov-Hinze scale, $d^+_H$, which can be computed as \citep{Perlekar2012,Roccon2017,Soligo2019c}:
\begin{equation}
d_H^+= 0.725 \left(\frac{We}{Re_\tau} \right) ^{-3/5} | \epsilon_c|^{-2/5}\, ,
\end{equation}
where $\epsilon_c$ is the turbulent dissipation, here evaluated at the channel center where most of the drops collect because of their deformability \citep{Lu2007,soligo2020effect,mangani2022influence}.
The Kolmogorov-Hinze scale identifies the critical diameter below which drop breakage is unlikely to occur \citep{kolmogorov1941local,Hinze1955}.
Separated by the Kolmogorov-Hinze scale, we observe the emergence of two different regimes.
%
%
For drops smaller than the Kolmogorov-Hinze scale, we find the coalescence-dominated regime (left, gray area), in which drops, which  are smaller than the critical scale, are generally not prone to break.
%
%
For drops larger than the Kolmogorov-Hinze scale, we find the breakup-dominated regime (right, white area) in which drops breakup is likely to happen.
Each regime is characterized by a specific scaling law, which describes the behavior of the drops number density  as a function of the drop size \citep{garrett2000,deane2002scale,chan2021a}: 
$PDF \sim {d_{eq}^+}^{-3/2}$, below Kolmogorov-Hinze scale,   and $PDF \sim {d_{eq}^+}^{-10/3}$ above it. The two scalings are represented by dot-dashed  lines in figure~\ref{dsd}.

We note that, for equivalent diameters   above the Hinze scale, our results follow quite well the theoretical scaling law and match the drops/bubbles size distributions obtained in literature considering similar flow instances \citep{deike2016,di2022coherent,soligo2021,deike2022mass,crialesi2023}.
Below the Hinze scale, for equivalent diameters in the range  $25<{d_{eq}^+}<{d_{H}^+}$ our results match reasonably well the theoretical scaling law. For equivalent diameters ${d_{eq}^+}<25~w.u.$,  we observe an underestimation of the DSD compared to the proposed scaling. 
This is linked to the grid resolution, and in particular to the problem in  describing very small drops \citep{soligo2021}.

\subsection{Mean temperature of drops and carrier fluid}

\begin{figure}
\begin{picture}(300,215)
\put(25,0){\includegraphics[width=0.8\columnwidth]{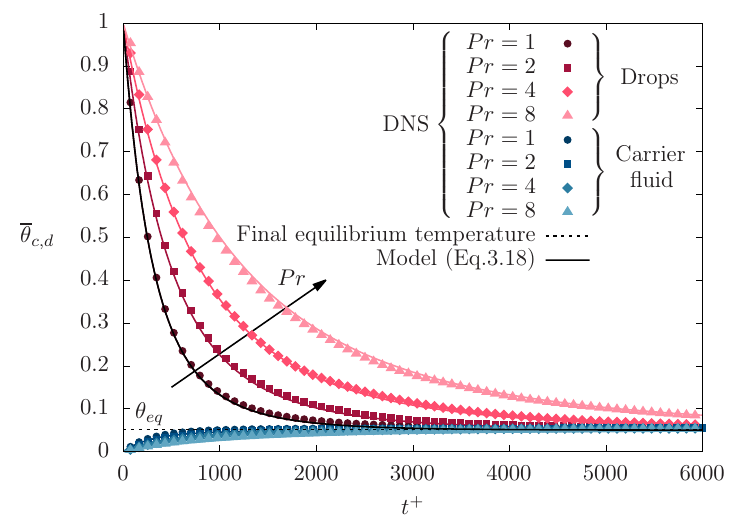}}
\end{picture}
\caption{Time evolution of the mean temperature of drops (violet to pink colors, different symbols) and carrier fluid (blue to cyan colors, different symbols) for the different Prandtl numbers considered.
DNS results are reported with full circles while the predictions obtained from the model are reported with continuous lines.
The equilibrium temperature of the system, $\theta_{eq}$,  is reported with a horizontal dashed line.
In general, it can be observed that by increasing the Prandtl number (corresponding to a decrease of the thermal diffusivity), the heat transfer process between the two phases slows down and more time is required to achieve the equilibrium condition.}
\label{meant}
\end{figure}

We now focus on the average temperature of the drops  and of the carrier fluid.
%
%
We consider the ensemble of all drops as one phase, and the carrier fluid as the other phase 
(using the value of the phase field as a phase discriminator), and we compute the average temperature for each phase.
%
%
The evolution in time of the drops and carrier fluid temperature, ${\overline{\theta}}_d$ and ${\overline{\theta}}_c$ respectively,  is shown 
in figure~\ref{meant}, for the different values of $Pr$. 
Together with the results obtained by current direct numerical simulations, filled symbols, in figure~\ref{meant} we also show the predictions obtained by a simplified phenomenological model (solid lines), the details of which will be described and discussed later (see section \ref{model_sec}).
We start considering the DNS results only.
%
As expected, we observe that the average temperature of the drops (violet to pink symbols) decreases in time,  while the average temperature of the carrier fluid (blue to cyan symbols) increases in time, until the thermodynamic equilibrium, at which both phases have the same temperature, is asymptotically reached.
%
For this reason, simulations have been run long enough  for the  average temperature of both phases  to be sufficiently close to the equilibrium temperature. 
In particular, we stopped the simulations at $t^+ \simeq 6000$, when the condition
\begin{equation}
\frac{(\overline{\theta}_d - \theta_{eq})}{(\theta_{d,0} -\theta_{eq})} \leq 0.05\, ,
\end{equation}
 with $\theta_{d,0}$ the initial temperature of the drops,  is satisfied by all simulations.
The equilibrium temperature, $\theta_{eq}$, can be easily estimated a-priori: since the two walls are adiabatic, and the homogeneous directions periodic, the energy of the system is conserved over time.
The energy balance can be written as:
\begin{equation}
m_c^* c_p^*  \theta_{d,0}^* +  m_d^* c_p^* \theta^*_{c,0}=  (m_d^* + m_c^*) c_p^* \theta^*_{eq}\, ,
\end{equation}
where $m_c^*$ and $m_d^*$ is the mass of the carrier fluid and of the drops; $\theta_{d,0}^*$ and  $\theta_{c,0}^*$ the physical value of the initial temperatures of the two phases and $c_p^*$ the specific heat capacity.
Considering that the two phases have equal density and specific heat capacity, we obtain: 
\begin{equation}
V^*_c c^*_p  \theta^*_{d,0} +  V^*_d c^*_p \theta^*_{c,0}=  (V^*_d + V^*_c) c^*_p \theta^*_{eq} \, .
\end{equation}
Recalling the definition of volume fraction, $\Phi=V_d/(V_d+ V_c)$, and making the equation dimensionless, we finally get:
\begin{equation}
\theta_{eq}=  \theta_{c,0}(1-\Phi)  +  \theta_{d,0} \Phi \ ,
\end{equation}
which is represented by the horizontal dashed line in Figure~\ref{meant}. 
%
Figure~\ref{meant} provides also a clear indication that the higher the Prandtl number, the larger the time it takes for the system to reach the equilibrium temperature, $\theta_{eq}$.
The trend can be observed for both the drops and carrier fluid, as the two phases are mutually coupled (the heat released from the drops is adsorbed by the carrier fluid).
This result confirms our previous qualitative observations, see figure~\ref{quali} and discussion therein, that a large $Pr$ (small thermal diffusivity) reduces the heat released by the drops.
%
%
It is also interesting to observe that the behavior of the mean temperature of the two phases appears self-similar at the different $Pr$. 
%

\subsection{A phenomenological model for heat transfer rates in droplet laden flows}
\label{model_sec}




In an effort to provide a possible  interpretation of the previous results --  and in particular to  explain the average temperature behavior shown in  figure~\ref{meant} --,  we develop a simple physically-sound model of the heat transfer process in droplet-laden turbulence.
We start by considering the heat transfer mechanisms from a single drop of diameter $d^*$ to the surrounding fluid:
\begin{equation}
 m^*_d c_p^*  \frac{\partial \theta^*_d}{\partial t^*}= \mathcal{H}^* A_d^* \left ( \theta_c^* - \theta_d^* \right )\, ,
 \label{ht1}
\end{equation}
where $m_d^*$,  $A_d^*$, and $c_p^*$  are the mass, external surface, and specific heat of the drop, $\mathcal{H}^*$ is the heat transfer coefficient, while  $\theta_d^*$ and $\theta_c^*$  are the drops and carrier fluid temperature.
The heat transfer coefficient can be estimated as the ratio between the thermal conductivity of the external fluid, $\lambda^*$,  and a reference length scale, here represented by the thermal boundary layer thickness $\delta_t^*$: 
\begin{equation}
\mathcal{H}^* \sim \lambda^* / \delta_t^*\, .
\end{equation}
With this assumption, and recalling that $\rho^*=\rho^*_c=\rho^*_d$, equation \ref{ht1} becomes:
\begin{equation}
\frac{\partial \theta_d^*}{\partial t^*}= \frac{6 }{ Pr } 
\frac{\nu^* }{ d^* \delta_t^*   }
\left( \theta_c^* - \theta_d^* \right )\, .
\end{equation}
%
%
Reportedly \citep[][218]{schlichting2016boundary},  the thermal boundary layer thickness, $\delta_t^*$, can be expressed as $\delta_t^*=\delta^* Pr^{-\alpha}$ where $\delta^*$ is the momentum boundary layer thickness, and $\alpha$ is an exponent that depends on the  flow condition in the proximity of the boundary where the boundary layer evolves.
In particular, the exponent $\alpha$ ranges from $\alpha=1/3$ for no slip conditions, usually assumed for solid particles, to $\alpha=1/2$, usually assumed for clean gas bubbles. For an in-depth discussion on the topic, we refer the reader to  appendix \ref{app_bl}.
As a consequence, the heat transfer rate observed from drops/bubbles is expected to be larger than that observed from solid particles,  since the no-slip boundary condition generally weakens the flow motion near the interface  \citep{levich1962,herlina2016isotropic,bird2002transport}.
%
We can now rewrite the  equation of the model in dimensionless form,  using the initial drop-to-carrier fluid temperature difference $\Delta \theta^*= \theta^*_{d,0} - \theta^*_{c,0}$ as reference temperature, and $\nu^*/u_{\tau}^{*2}$ as reference time:
\begin{equation}
\frac{\partial \theta_d}{\partial t^+}=   6 Re_{\delta}^{-1} Pr ^{-1+\alpha} \left ( {d^+} \right)^{-1}   \left( \theta_c - \theta_d \right )\, ,
\label{ht3}
\end{equation}
where $d^+$ is the drop diameter in wall units, while $Re_{\delta}=u_{\tau}^* \delta^* /\nu^*$ is the Reynolds number based on the boundary layer thickness (which can be assumed constant among the different cases). 
Equation \ref{ht3} can be rewritten as: 
\begin{equation}
\frac{\partial \theta_d}{\partial t^+}=    \mathcal{C} Pr ^{-1+\alpha} \left ( {d^+} \right)^{-1}   \left( \theta_c - \theta_d \right )\, ,
\label{ht4}
\end{equation}
where $\mathcal{C }$ is a constant whose value depends only on the flow structure, i.e. on  $\Rey_{\delta}$.
%
Equation \ref{ht4} describes the heat released by a single drop of dimensionless diameter $d^+$. 
Assuming now that the turbulent flow is laden with drops of different diameters the general equation describing the heat released by the $i$-th drop of diameter $d_i^+$ becomes:
\begin{equation}
\frac{\partial \theta_{d,i}}{\partial t^+}=    \mathcal{C}  Pr ^{-1+\alpha} \left ( {d^+_i} \right)^{-1} \left( \theta_c - \theta_d \right )\,=\mathcal{F}_i\, ,
\label{ht5}
\end{equation}
where $\mathcal{F}_i$ is the lumped-parameters representation of the right-hand side of the temperature evolution equation for the $i$-th drop.
As widely observed in literature \citep{deane2002scale,Soligo2019b}, and also confirmed by the present study (figure~\ref{dsd}), 
 we can hypothesize an equilibrium drops-size-distribution (DSD) by which the number density of drops scales as ${d^+}^{-3/2}$, in the sub-Hinze range of diameters ($10<d^+<110$), and as ${d^+}^{-10/3}$ in the super-Hinze range of diameters ($110<d^+<240$). 
With this assumption, and considering 7 classes of drops diameter for the sub-Hinze range, and 4 classes for the super-Hinze range, we can integrate equation \ref{ht5} to obtain the time evolution of the temperature of each drop in time: 
\begin{equation}
\theta_{d,i}^{n+1}= \theta_{d,i}^{n}+ \Delta t^+ \mathcal{F}_i\, .
\label{ht8}
\end{equation} 
From a weighted averaged of the temperature  (based on the number of drops in each class, as per the theoretical DSD), we  obtain the average  temperature of the drops, $\overline{\theta}_d$.

To obtain the mean temperature  of the carrier fluid, we  consider that (adiabatic condition at the walls), the heat released by the drops is entirely absorbed by the carrier fluid. 
The heat released by the drops with a certain diameter $d_i^*$, can be computed as:
\begin{equation}
Q_i^*=m_d^* c_p^* \frac{\partial \theta_d}{\partial t^+}N^*_d(i)\, ,
\label{ht6}
\end{equation}
where $N^*_d(i)$ is the number of drops for that specific diameter (as per the DSD).
The overall heat released by all drops can be calculated as the summation over all different classes of diameters:
\begin{equation}
Q^*_{tot}=\sum_{i=1}^{N_{c}}  Q_i^*\, ,
\label{ht7}
\end{equation}
where $N_c$ is the employed number of classes.
Finally, the  mean temperature of the carrier fluid is 
\begin{equation}
\overline{\theta}_c^{*,n+1}= \theta_c^{*,n}+ \Delta t^+ \frac{Q_{tot}^*}{m_c^* c_p^*}\, .
\label{ht9}
\end{equation}
In dimensionless form (dividing by the initial drop-to-carrier fluid temperature $\Delta \theta^*$) equation \ref {ht9}  becomes:
\begin{equation}
\overline{\theta}_c^{n+1}= \theta_c^{n}+ \Delta t^+ Q_{tot}\, .
\end{equation}

\begin{figure}
\begin{picture}(300,215)
\put(25,0){\includegraphics[width=0.8\columnwidth]{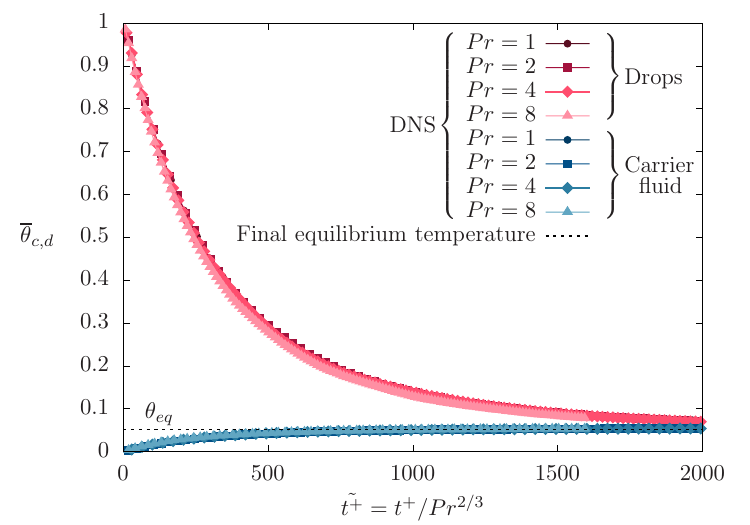}}
\end{picture}
\caption{Time evolution of the mean temperature of drops (violet to pink colors) and carrier fluid (blue to cyan colors) for the different Prandtl numbers considered obtained from DNS and reported against the dimensionless time $\tilde{t}^+= t^+/Pr^{2/3}$. The equilibrium temperature of the system, $\theta_{eq}$,  is reported with a horizontal dashed line. 
The DNS results reported over the new dimensionless time nicely collapse on top of each other, highlighting the self-similarity of the $\overline{\theta}_{c,d}$  profiles.}
\label{rescaling}
\end{figure}

The results of the model are shown in figure~\ref{meant}.
Interstingly,  under the simplified hypothesis of the model (chiefly, spherical shape of the drops, constant drop-size-distribution evaluated at the equilibrium), we observe that the behavior of the mean temperature is very well captured by the model (represented by the solid lines in figure~\ref{meant}) 
\begin{equation}
\frac{\partial \theta_d}{\partial t^+}=   \mathcal{C} Pr ^{-2/3} \left ( {d^+} \right)^{-1}   \left( \theta_c - \theta_d \right )\, ,
\label{ht10}
\end{equation}
i.e. when $\alpha=1/3$ -- typical of  boundary layers  around solid objects (i.e. solid particles).
Reasons for this might be the presence of wakes/sheltering effects between  drops, but also  the fact that drops are strongly advected by the mean flow, and the flow condition at the drop surface can be different from the slip one, and is in general not of  simple evaluation.
%
%
Given the relationship  $\partial \theta_d/\partial t \sim Pr^{-2/3}$ postulated by the model (equation \ref{ht10}), which provides results in very good agreement with the numerical ones, it seems reasonable to rescale the  time variable as:
\begin{equation}
\tilde t^+=\frac{t^+}{Pr^{(1-\alpha)}}=\frac{t^+}{Pr^{(2/3)}}\, .
\label{ttilde}
\end{equation}
%
%
A representation of the DNS results in terms of the rescaled time, equation~\ref{ttilde}, is shown in 
figure~\ref{rescaling}. 
%
We observe a nice collapse of the two sets of curves -- drops and carrier fluid (red and blue) --  for the different  values of $Pr$, which clearly demonstrates the self-similar behavior of $\overline{\theta}$.
%
%
For this reason,  the rescaling of time $\tilde t^+={t^+}/Pr^{2/3}$, will be also used in the following.
%

\subsection{Heat transfer from particles and drops/bubbles}
\label{appb}

It is now important to discuss the behavior of the  heat transfer coefficient (and its dimensionless counterpart, the Nusselt number $Nu$),  also in the context of available literature results. 
Naturally, similar considerations can be  made to evaluate the mass transfer coefficient, in particular at liquid/gas interfaces  \citep{levich1962,bird2002transport}.

For solid particles, a balance between the convective time scale  near the surface, and the diffusion time scale, gives an heat transfer coefficient \citep{krishnamurthy2018heat}:
\begin{equation}
\mathcal{H}^* \propto Pr^{-2/3}\, ,
\end{equation}
and the corresponding  Nusselt number:
\begin{equation}
Nu   \propto Re^{\beta} Pr^{1/3}\,, 
\end{equation}
where $\beta$ is an exponent that depends on the flow conditions and links the boundary layer thickness to the  particle Reynolds number. Usually, $\beta=1/3$ for small Reynolds numbers \citep{krishnamurthy2018heat} while $\beta=1/2$ for large Reynolds numbers \citep{ranz1952evaporation,whitaker1972forced,michaelides2003hydrodynamic}.

Using similar arguments (balance between  convective and diffusion time scales), but considering now that at the surface of a drop/bubble a slip velocity,  and therefore a certain degree of advection, can be observed \citep{levich1962,bird2002transport,herlina2016isotropic}, the heat transfer coefficient is found to scale as:
\begin{equation}
\mathcal{H}^* \propto Pr^{-1/2}\, , 
\end{equation}
and the corresponding Nusselt number as:
\begin{equation}
Nu   \propto Re^{\beta} Pr^{1/2}\, ,
\end{equation}
where also in this case the exponent $\beta$ does  depend on the considered Reynolds number. 
Two regimes are usually defined \citep{theofanous1976turbulent}: a low Reynolds number regime, for which  $\beta=1/2$, and a high Reynolds number regime,  for which $\beta=3/4$.
An alternative approach, which gives similar predictions, is to use the penetration theory of  \citet{higbie1935rate}, in which turbulent fluctuations are invoked to estimate a flow exposure (or contact) time, to compute the heat/mass transfer coefficient. Such approach has been widely used in  bubble-laden flows \citep{colombet2011experimental,herlina2014direct,herlina2016isotropic,farsoiya2021bubble}.

We can now evaluate the heat transfer coefficient  from our  DNS at different $Pr$, and compare it to the proposed scaling laws.
Note that the heat transfer coefficient is obtained as:
\begin{equation}
\mathcal{H}= \frac{(\overline{\theta}_d^{n+1} - \overline{\theta}_d^{n})}{A \Delta t (\overline{\theta}_d^{n+1/2} - \overline{\theta}_c^{n+1/2})}\, ,
\label{coef}
\end{equation}
where the numerator represents the temperature difference of the drops  between the time steps $n$ and $n+1$, while the denominator represents  the temperature difference between the drop and the carrier fluid evaluated halfway in time between step $n$ and $n+1$ (i.e. at $n+1/2$). The quantity $A$ is the total interfacial area between drops and carrier fluid, while  $\Delta t$ is the time step used to evaluate the heat transfer.
Here, we have evaluated the heat transfer coefficient taking the heat released by the drops as  reference; an equivalent result, but with  opposite sign, can be obtained using the heat absorbed by the carrier fluid as reference, and taking into account the different volume fraction of the two phases.

\begin{figure}
\begin{picture}(300,180)
\put(40,-20){\includegraphics[width=0.8\columnwidth]{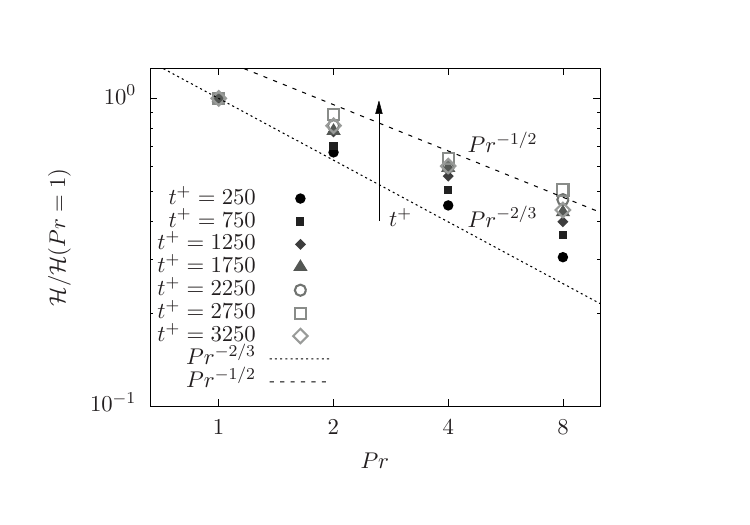}}
\end{picture}
\caption{Time behavior of the dimensionless heat transfer coefficient for the different Prandtl numbers considered.
Heat transfer coefficients are reported normalized by the value of the heat transfer coefficient obtained for $Pr=1$ (at the same time instant).
In this way, results obtained at different time instants can be conveniently compared.
The two scaling laws that refer to $\alpha=2/3$ and $\alpha=1/2$ are also reported as references. }
\label{scaling}
\end{figure}

The dimensionless heat transfer coefficient, equation~\ref{coef}, is shown as a function of $Pr$, and at different time instants, in  figure~\ref{scaling}.
For a better comparison, the results are normalized by the value of the heat transfer coefficient for $Pr=1$.
%
%
The two reference scaling laws, $\mathcal{H}\sim Pr^{-2/3}$ obtained for $\alpha=1/3$ and  $\mathcal{H}\sim Pr^{-1/2}$ obtained for $\alpha=1/2$  are also shown by a dotted and a dashed line.
We note that at the beginning of the simulations (see for example $t^+=250$), the heat transfer coefficient is close to $\mathcal{H}\sim Pr^{-2/3}$, while at later times it tends towards $\mathcal{H}\sim Pr^{-1/2}$, hence approaching the scaling law proposed for heat/mass transfer in gas-liquid flows  \citep{levich1962,magnaudet2000motion,bird2002transport,herlina2014direct,herlina2016isotropic,colombet2018single,farsoiya2021bubble}.
%
%
%

%
A possible explanation is that, as time advances, the shape of the drops becomes  complex, and coalescence/breakups more frequent, thus inducing an overall surface decrease that is associated to an heat transfer increase.
This reflects  into an heat transfer process that is slower at the beginning, 
$\mathcal{H}^* \sim Pr^{-2/3}$,
and faster at later times
$\mathcal{H}^* \sim Pr^{-1/2}$.

\subsection{Influence of the drop size on the average drop temperature}

\begin{figure}
\setlength{\unitlength}{0.0025\columnwidth}
\begin{picture}(400,300)
\put(0,280){$(a)$}
\put(206,280){$(b)$}
\put(0,138){$(c)$}
\put(206,138){$(d)$}
\put(37,280){$Pr=1$}
\put(227,280){$Pr=2$}
\put(37,138){$Pr=4$}
\put(227,138){$Pr=8$}
\put(-15,145){\includegraphics[width=0.58\columnwidth]{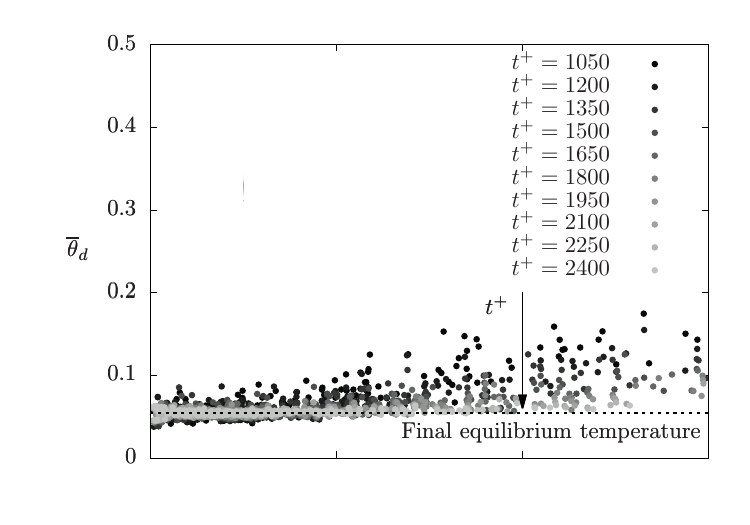}}
\put(175,145){\includegraphics[width=0.58\columnwidth]{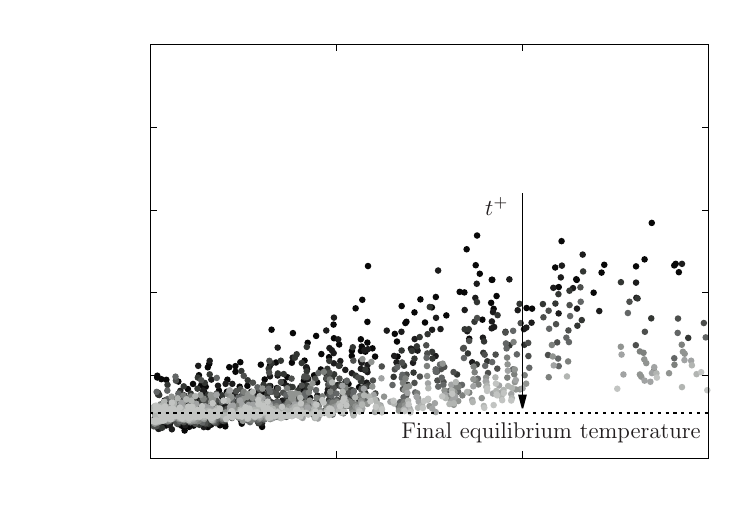}}
\put(-15,0){\includegraphics[width=0.58\columnwidth]{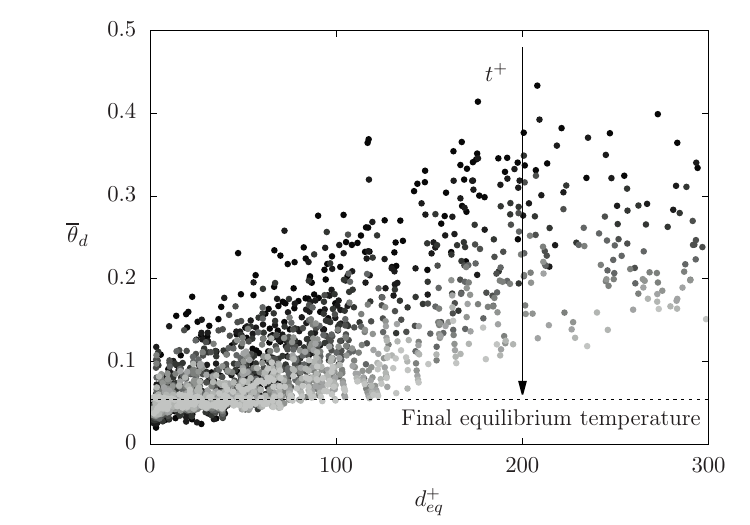}}
\put(175,0){\includegraphics[width=0.58\columnwidth]{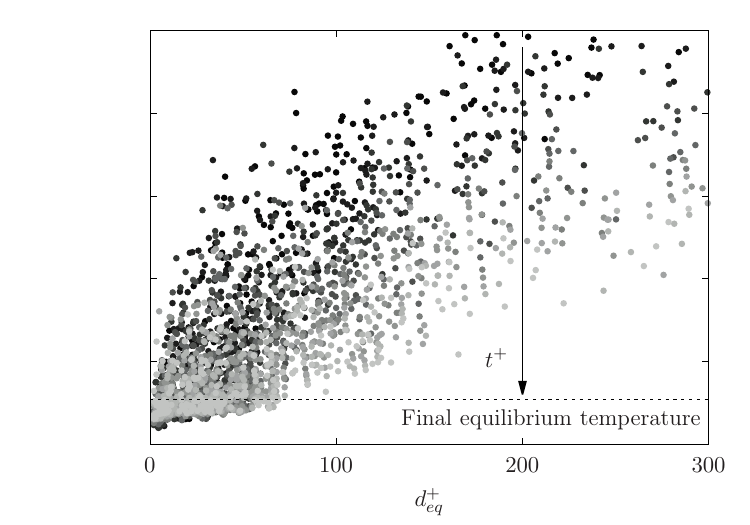}}
\put(35,230){\includegraphics[width=0.27\columnwidth]{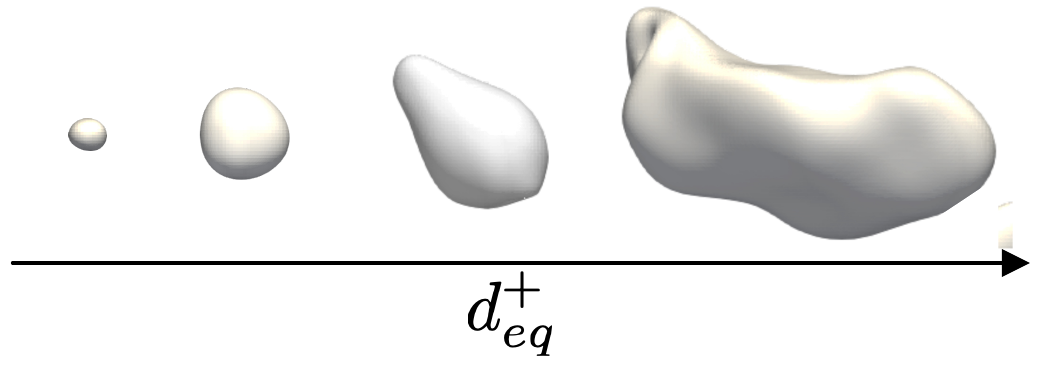}}
\end{picture}
\caption{Scatter plot of the drop equivalent diameter $d_{eq}^+$ against the drop average temperature $\overline{\theta}_d$.
Each dot represents a different drop while its color (black to gray colormap) identifies different times, from $t^+=1050$ (black) up to $t^+=2400$ (gray).
Each panel refers to a different Prandtl number.
A sketch showing drops of different equivalent diameters is reported in the upper part of panel $a$. 
As can be appreciated, larger drops have larger relaxation times, and thus a longer transient is required to reach thermal equilibrium.
This effect becomes more pronounced as the Prandtl number is increased and thus the overall heat transfer process slows down (smaller heat transfer coefficient).}
\label{scatter}
\end{figure}

In the previous sections, we have studied the behavior of the mean temperature field of the drops and of the carrier fluid considered    as   single entities.
However, while this description is perfectly reasonable for the carrier fluid -- which  can be considered a continuum -- it can be questionable for the drops, which are not a continuum phase by nature.
%
%
%
We now take the dispersed nature of the drops into account and we evaluate, for each drop, the equivalent diameter and the corresponding mean temperature.
%

This is sketched in  figure~\ref{scatter},
where the average temperature of each drop (represented by a dot) is shown as a function of its equivalent diameter, at different time instants  (between $t^+=1050$ and $t^+=2400$). 
Each panel refers to a different Prandtl number.
%
%
%
Note that, at $t^+=2400$,  the case $Pr=1$ has almost reached the thermodynamic equilibrium (figure~\ref{meant}).
It is clearly visible that, regardless of the  considered time, small drops have an average temperature close to the equilibrium one. This is particularly visible at smaller Prandtl numbers, i.e. when  heat transport is faster, but it can be observed also at larger $Pr$.
In contrast, the average temperature of larger drops is larger.
%
Hence, the  average temperature of the drops seems directly proportional to the drop size, as can be argued considering that the heat released by the drop, and hence its temperature reduction, is $\partial \theta_d/\partial t\propto d^{-1}$ (equation \ref{ht9}). 
It is therefore not surprising that the scatter plot  at a given time instant is characterized by dots distributing   in stripes-like fashion,  with  slope that decreases with time.
This behavior is  observed at all $Pr$,  although the  range of drops temperature (y axis) at small $Pr$ is definitely narrower (because of their larger heat loss) compared to that at large $Pr$.
%
%
It is also interesting to note -- in particular  at $Pr=4$ and $Pr=8$ (panels $c$ and $d$) -- the presence of drops with a temperature smaller than the equilibrium one (dots falling below the horizontal line that marks  the equilibrium temperature). 
We can link this behavior to the small relaxation time of small drops that therefore adapt quickly to the local  temperature of the carrier fluid, which can be smaller than the equilibrium one for two main   reasons.
First, at the early stages of the simulations, and at high Prandtl numbers, the temperature of the carrier fluid is lower than the equilibrium one.
Second, temperature fluctuations (of both negative and positive sign) are present also in the carrier fluid.
These fluctuations, in the form of hot/cold striations are more likely observed at large $Pr$ (see the  striation-like structures  at $Pr=8$  in figure~\ref{quali}$d$).

\subsection{Temperature distribution inside the drops}

\begin{figure}
\setlength{\unitlength}{0.0025\columnwidth}
\begin{picture}(400,320)
\put(15,290){$(a)$}
\put(208,290){$(b)$}
\put(15,132){$(c)$}
\put(208,132){$(d)$}
\put(48,290){$t^+=600$}
\put(228,290){$t^+=1500$}
\put(48,132){$\tilde t^+=600$}
\put(228,132){$\tilde t^+=1500$}
\put(170,308){Dimensionless time $t^+$}
\put(160,148){Dimensionless time $\tilde{t}^+=t^+/Pr^{2/3}$}
\put(0,160){\includegraphics[width=0.54\columnwidth]{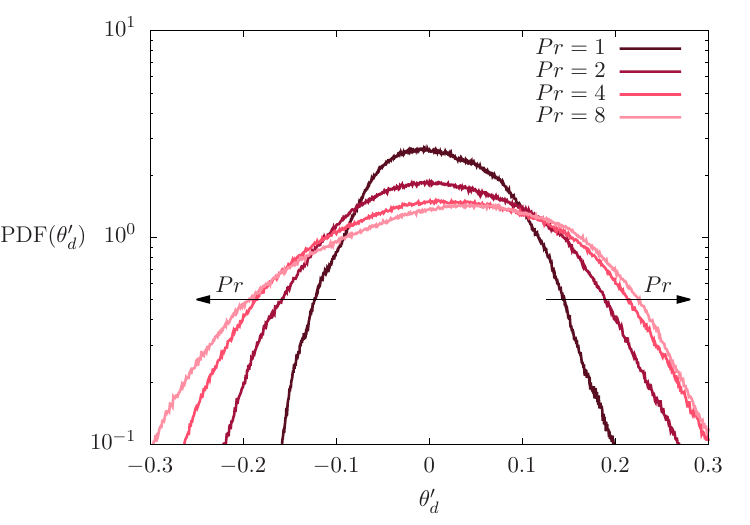}}
\put(180,160){\includegraphics[width=0.54\columnwidth]{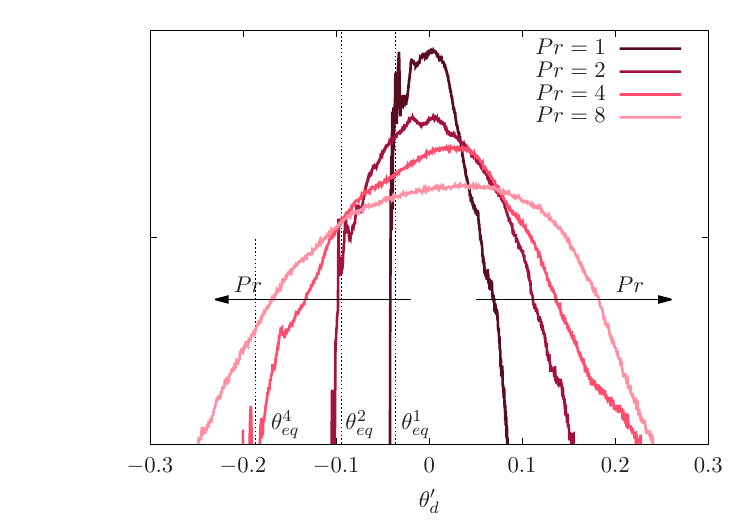}}
\put(0,0){\includegraphics[width=0.54\columnwidth]{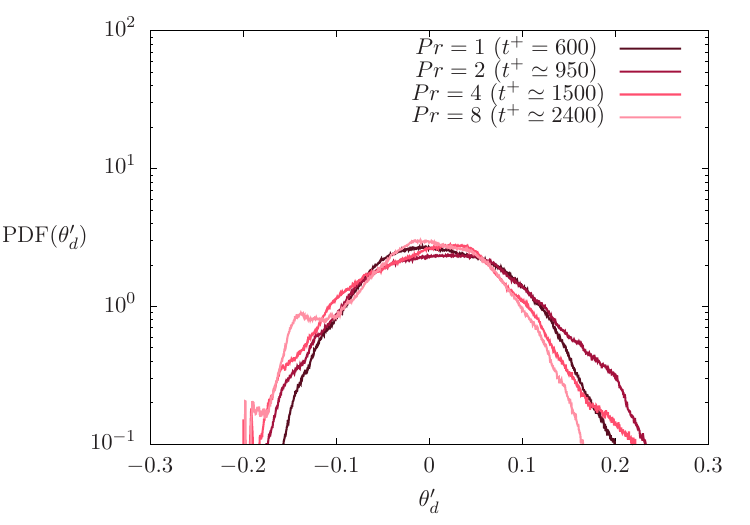}}
\put(180,0){\includegraphics[width=0.54\columnwidth]{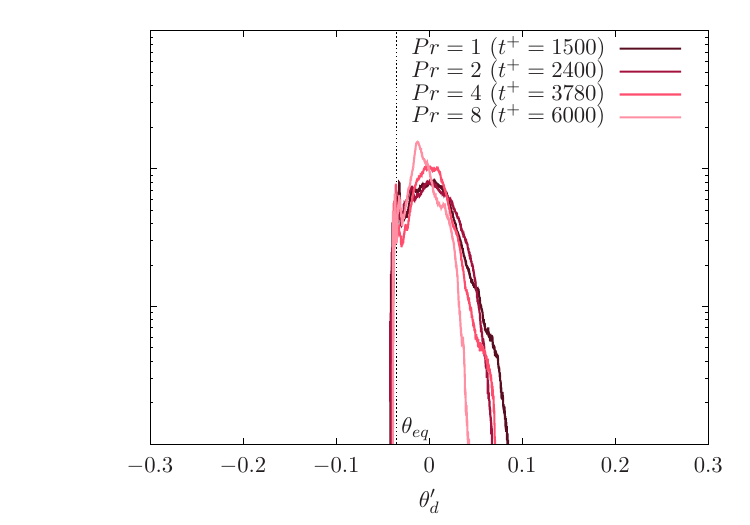}}
\put(45,245){\includegraphics[width=0.10\columnwidth]{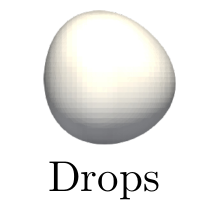}}
\put(45,87){\includegraphics[width=0.10\columnwidth]{../figures/dphase}}
\put(225,245){\includegraphics[width=0.10\columnwidth]{../figures/dphase}}
\put(225,87){\includegraphics[width=0.10\columnwidth]{../figures/dphase}}
\end{picture}
\caption{Probability density function (PDF) of the temperature fluctuations, $\theta'_d = \theta_d - \overline{\theta}_d$ inside the drops.
Each case is reported with a different color (violet to light pink) depending on the Prandtl number.
The first row shows the PDFs obtained at two different time instants: $t^+=600$ (panel $a$) and $t^+=1500$ (panel $b$).
The second row shows the PDFs obtained at two rescaled time instants $\tilde{t}^+=600$ (panel $c$) and $\tilde{t}^+=1500$ (panel $d$), where the rescaled time is computed as $\tilde{t}^+=t^+/Pr^{2/3}$. 
For panels $c$-$d$, the corresponding $t^+$ is reported between brackets.}
\label{pdft}
\end{figure}

In many applications, in particular to evaluate mixing efficiency and flow homogeneity, not only the average temperature of drops is important,  but also its space and time distribution inside the drops. 
%
%
%
To understand it, we now look  at  the PDF of the temperature fluctuations inside the drops,
\begin{equation}
\theta^\prime_d =\theta_d - {\overline\theta}_d\, ,
\end{equation}
where ${\theta}_d$ is the local  temperature inside the drop, and ${\overline\theta}_d$ is the average temperature of all drops at a certain time (as per figure \ref{meant}). Results are shown in figure~\ref{pdft}.
The first row of figure~\ref{pdft} shows the probability density function of $\theta^\prime_d$ at different $Pr$, and at two different time instants: $t^+=600$ (panel $a$, left) and $t^+=1500$ (panel $b$, right).
The second row of figure~\ref{pdft} shows the PDFs obtained at two different rescaled time instants, $\tilde{t}^+=t^+/ Pr^{2/3}$: $\tilde{t}^+=600$ (panel $c$, left) and $\tilde{t}^+=1500$ (panel $d$, right).

Considering first figure~\ref{pdft}$a$ ($t^+=600$), we notice that all PDFs have a rather regular shape, characterized by the presence of  both positive and negative fluctuations (with respect to the average temperature), with a slight asymmetry towards the positive ones (positive fluctuations are more likely  observed).
%
%
A comparison between the curves obtained at different $Pr$ shows  that  the range of temperature fluctuations is wider at larger $Pr$. 
This  is due to the small thermal diffusivity at large $Pr$, which allows temperature fluctuations  in the drop to survive much longer before they are damped and spread by diffusion.
Naturally, at later times (figure~\ref{pdft}$b$, $t^+=1500$), the range of temperature fluctuations reduces.
Indeed, as heat is transferred from the drops to the carrier fluid, the maximum temperature of drops reduces, and so does the range of temperature fluctuations inside the drop.
This trend is more pronounced for negative fluctuations, as the minimum temperature inside the drops is somehow bounded by the temperature of the carrier fluid (which increases only  a little, from $\overline{\theta}_{c,0}=0$ to $\theta_{eq}=0.054$, during the simulation).
This latter observation is  visible in the shape of the PDFs  at $Pr=1, 2$ and $4$, since the system is closer to the thermal equilibrium at this time instant (the thermal equilibrium is identified  in panel $b$ by a vertical dashed line and marked with a label, $\theta_{eq}^{Pr}$): a sharp drop of the PDF, which does not significantly trespass the $\theta_{eq}^{Pr}$ limit, is observed.
%
%
In contrast, positive temperature  fluctuations are subject to relatively weaker constraints (they are only bounded by the maximum  initial temperature of the drops). This results into a PDF that gets  asymmetric, positively-skewed.
It is also interesting to observe the development of a pronounced peak about the equilibrium temperature $\theta_{eq}^{Pr}$, which corresponds to the presence of small drops (generated by breakages events) that -- given  their small thermal relaxation time and heat capacity -- almost immediately adapt to the equilibrium temperature  (see also figure~\ref{time}$d$,$f$).

However, a discussion on the temperature fluctuations, captured from  flows at different $Pr$ and after the same time $t^+$ from the initial condition, could be misleading, because it puts in contrast flows at different thermal states (i.e. different average temperature, and different temperature gradients, see figure \ref{meant}).
To filter out this effect, we compute the PDFs of the temperature fluctuations at the same rescaled time instants $\tilde{t}^+=t^+/Pr^{2/3}$.
By doing this, all cases can be considered at similar thermal conditions  (see also figure~\ref{rescaling}). The resulting PDFs, at  $\tilde{t}^+=600$ and $\tilde{t}^+=1500$, are shown in figure~\ref{pdft}$c$-$d$. Note that, for the sake of clarity, the corresponding  $t^+$, which is different from case to case, is reported between brackets in the legend.
In the rescaled time units, the collapse between the different curves is quite nice. 
The slight difference between the curves is due to the fact that, although the system is at the same thermal state (same $\tilde{t}^+$), it is at different flow state  (different $t^+$), i.e. the instantaneous drop size distributions are different.  
This gives the slightly larger negative fluctuations at larger $Pr$ (which, being at a later stage, is characterized by the presence of smaller and colder drops), and slightly larger positive  fluctuations at smaller $Pr$ (which, being at an earlier flow state, is characterized by the presence of larger and warmer drops).

%

From a closer look at   figure~\ref{pdft}$d$ ($\tilde{t}^+ =1500$), we note very clearly the 
 constraint set by the thermal equilibrium condition: the PDF cannot significantly trespass the limit represented by $\theta_{eq}$ (vertical dashed line), which is very similar for all $Pr$, given the similar thermal state.
Also visible is the peak, already discussed in 
figure~\ref{pdft}$b$, that emerges very close to the equilibrium temperature $\theta_{eq}$, and which is due to  the presence of small drops that  adapt quickly to the local temperature of the carrier fluid.
As previously noticed in  figure~\ref{pdft}$c$, the higher probability of finding small drops at lower $Pr$ is also responsible for the narrowing of the PDF (reduction of positive temperature fluctuations).

\section{Conclusions}

In this work, we studied heat transfer in a turbulent channel flow laden with large and deformable drops.
The drops are initially warmer than the carrier fluid and as the simulations advance, heat is transferred from the drops to the carrier fluid.
Simulations considered a fixed value of the Reynolds number, $\Rey_\tau=300$, and Weber number, $We=3$, and analyze different Prandtl number values, from $Pr=1$ to $Pr=8$.
The Prandtl number is  changed by changing the thermal diffusivity.
The investigation is based on the direct numerical simulation of turbulent heat transfer, coupled with a phase-field method, used to describe interfacial phenomena. 
%
First, we focused on the  drops dynamics, observing that after an initial transient (up to $t^+=1000$), the drop size distribution (DSD) reaches a quasi-equilibrium condition where it follows the scaling ${d^+_{eq}}^{-3/2}$ in the coalescence-dominated regime and ${d^+_{eq}}^{-10/3}$ in the breakage-dominated regime. The threshold between the coalescence-dominated and the breakage-dominate regimes  is represented by the Kolmogorov-Hinze scale.
%
%
Then, we characterize the  behavior of the
average temperature of the drops and of the carrier fluid: as expected, the average  temperature of drops decreases in time, while the average temperature of the carrier fluid increases  in time, until reaching the equilibrium condition of uniform temperature in the whole system.
We clearly observed that 
the higher the Prandtl number, the larger the time it takes for the system to reach the equilibrium temperature. Interestingly, the time behavior of the temperature profiles of both drops and carrier fluid is self-similar. 
%
Building on top of these numerical results, we developed a phenomenological model that can accurately reproduce the time evolution of the mean temperatures at all Prandtl numbers considered here.
This model gave us the opportunity to introduce a new self-similarity variable (time,  $\tilde{t}^+$) that accounts for the Prandtl number effect, and by which all results collapse on a single curve.
In addition, we also computed the heat transfer coefficient $\mathcal{H}$ (and its dimensionless counterpart, the Nusselt number $Nu$) and showed that it scales as $\mathcal{H}\sim Pr^{-2/3}$  (which corresponds to a Nusselt number scaling $Nu\sim Pr^{1/3}$) at the beginning of the simulation,   and tends to $\mathcal{H}\sim Pr^{-1/2}$  (or, alternatively, $Nu\sim Pr^{1/2}$) at later times. 
These different scalings are consistent with previous literature  predictions, and can be explained via the boundary layer theory (appendix \ref{app_bl}).
The effects of the Prandtl number on the temperature distribution inside the drops has been investigated.
We observe that by increasing the Prandtl number, the PDFs become wider and thus large temperature fluctuations are more likely to be observed.
Interestingly, when the PDFs are compared at a the same rescaled time $\tilde{t}^+$ (i.e. accounting for the Prandtl number effect), all  curves collapse on top of each other, with only minor differences possibly due to the different instantaneous drop size distribution.
The effect of the drop size was also discussed: small drops adapt faster to the equilibrium temperature,  thanks to their small heat capacity, compared to larger drops.
Finally, it must be  pointed out that, since the different phases of a multiphase flow  can have different thermophysical properties, Prandtl numbers can be also different from phase to phase. This aspect, which was not considered in the present work, will be the topic of a future study.
In addition,  in the present work we have assumed a constant and uniform surface tension. However, in many circumstances, surface tension does depend on temperature, therefore inducing thermocapillary effects. This will be also the subject of a future investigation. 
 

\backsection[Acknowledgements]{We acknowledge EURO-HPC JU for awarding us access to Discoverer@Sofiatech, Bulgaria (Project ID: EHPC-REG-2022R01-048) and LUMI-C@LUMI, Finland (Project ID: EHPC-EXT-2022E01-003) and ISCRA for awarding us access to Leonardo (Project ID: HP10BUJEO5).}

\backsection[Funding]{FM gratefully acknowledges funding from the MSCA-ITN-EID project COMETE (Project No. 813948) and AR gratefully acknowledges funding from PRIN 2017 - Advanced Computations \& Experiments for anisotropic particle transport in turbulent flows (ACE).}

\backsection[Declaration of interests]{The authors report no conflict of interest.}

\backsection[Data availability statement]{Data available on request from the authors.}

\backsection[Author ORCIDs]{
\\
\noindent
Francesca Mangani, https://orcid.org/0000-0001-7777-6665\\
Alessio Roccon, https://orcid.org/0000-0001-7618-7797 \\
Francesco Zonta, https://orcid.org/0000-0002-3849-315X \\
Alfredo Soldati, https://orcid.org/0000-0002-7515-7147}

\backsection[Author contributions]{FM performed the simulations.  FM and AR analysed the data. FCZ developed the model.
All authors contributed equally  in writing the paper.}

\appendix

\section{Effects of slip condition on the velocity and thermal boundary layer evolution}
\label{app_bl}

In this section, we derive and solve the equations that describe the evolution of the boundary layer   on a heated flat plate that is parallel to a constant unidirectional flow.

\begin{figure}
\begin{picture}(300,140)
\put(0,110){$(a)$}
\put(200,110){$(b)$}
\put(0,0){\includegraphics[width=1.0\columnwidth]{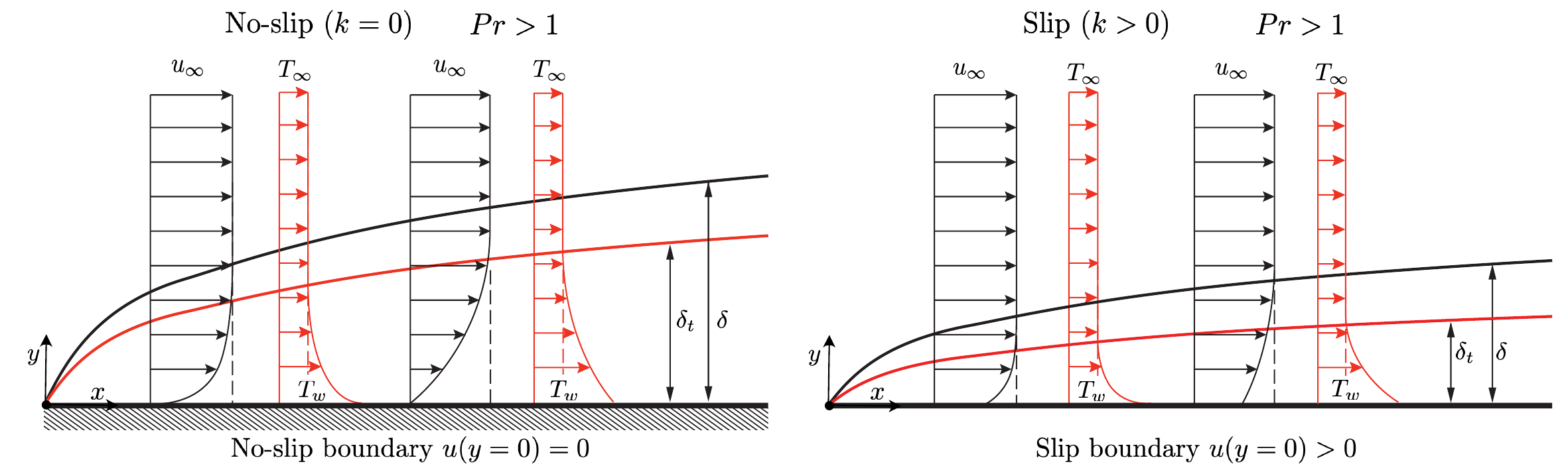}}
\end{picture}
\caption{Sketch of the momentum and thermal boundary layer dynamics on a flat plate characterized by a uniform temperature, $T_w$, larger than the free stream temperature, $T_\infty$. 
In panel~$a$ no-slip conditions are enforced at the wall (corresponding to a slip parameter $k=0$) while in panel $b$ partial slip is allowed at the wall.
The qualitative behavior of the momentum and thermal boundary layer thickness is also shown for the two cases.
Both panels refer to a super-unitary  Prandtl number. }
\label{bl}
\end{figure}

In addition to the standard description of the boundary layer, where no-slip conditions on the plate are considered \citep{prandtl1905uber,blasius1908grenzschichten}, here we consider also the effect of a slip velocity on the velocity and thermal boundary layers \citep{martin2006momentum,bhattacharyya2011steady,aziz2014steady}.
Following the standard approach \citep{schlichting2016boundary}, the continuity, Navier-Stokes and energy equations in 2D are:
\begin{equation}
\frac{\partial u}{\partial x} + \frac{\partial v}{\partial y}= 0\, ,
\end{equation}
\begin{equation}
u \frac{\partial u}{\partial x}+ v \frac{\partial u}{\partial y}= - \frac{1}{\rho}\frac{\partial p}{\partial x} +  \nu \frac{\partial^2 u}{\partial y^2} \, ,
\end{equation}
\begin{equation}
u \frac{\partial T}{\partial x}+ v \frac{\partial T}{\partial y}= a \frac{\partial^2 T}{\partial y^2} \, ,
\end{equation}
where $x$ is the direction parallel to the wall, and $y$ the direction normal to the wall, see figure~\ref{bl}.
The boundary conditions, accounting also for the slip velocity, read as:
\begin{equation}
u(x,y=0)=k \frac{\partial u}{\partial y} (x,y=0)\, ,
\label{bc1}
\end{equation}
\begin{equation}
v(x,y=0)=0\, ,
\end{equation}
\begin{equation}
u(x,y  \rightarrow + \infty)= u_\infty \, ,
\label{bc3}
\end{equation}
\begin{equation}
T(x,y=0)= T_w \, ,
\end{equation}
\begin{equation}
T(x,y  \rightarrow + \infty)= T_\infty \, ,
\end{equation}
where $k$ is a parameter that controls the amount of slip at the wall (no-slip for $k=0$,  up to free-slip for $k \rightarrow + \infty$), $u_\infty$ and $T_\infty$ are the free stream velocity and temperature, and $T_w$ is the constant temperature of the flat plate.

\begin{figure}
\begin{picture}(300,160)
\put(3,130){$(a)$}
\put(195,130){$(b)$}
\put(-5,0){\includegraphics[width=0.55\columnwidth]{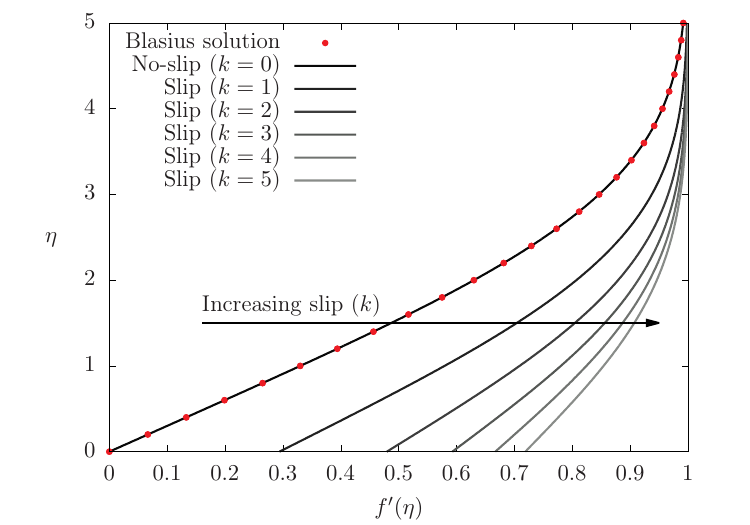}}
\put(188,0){\includegraphics[width=0.55\columnwidth]{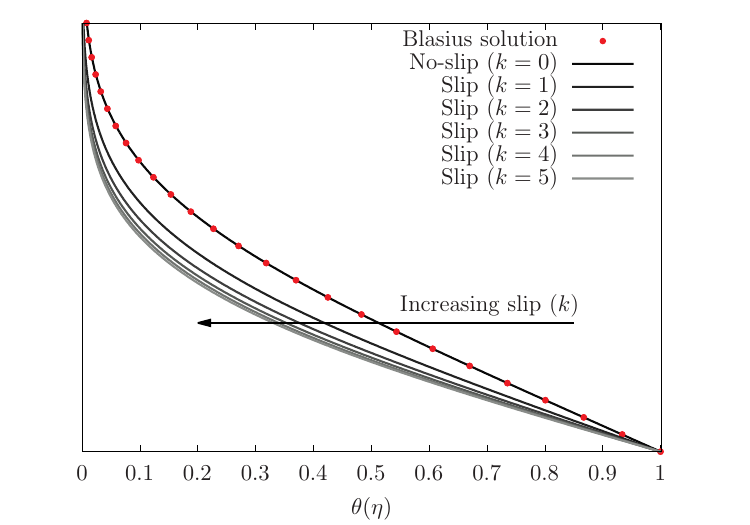}}
\end{picture}
\caption{Streamwise velocity (panel~$a$) and temperature profiles (panel $b$) obtained for different values of the slip parameter $k=0$.
Results are reported rotated by $90^\circ$ degrees for sake of better interpretation and are obtained considering $Pr=1$. 
For the no-slip case ($k=0$), the classical Blasius solution available in archival literature for the velocity, $f'$, and temperature, $\theta=1-f'$, is reported with red dots.
By increasing the slip parameter $k$, velocity at the wall location $\eta=0$ increases and larger temperature gradients are observed.}
\label{esempi}
\end{figure}

To solve the system of equations, we use the method of similarity transformation. First, we consider the continuity and Navier-Stokes equations.
Following \citet{blasius1908grenzschichten}, we introduce the following similarity transformation:
\begin{equation}
\eta=y \sqrt{\frac{u_{\infty}}{\nu x}}\, .
\end{equation}
We can define a dimensionless stream function, $f(\eta)$, which depends only on the variable $\eta$,
\begin{equation}
f(\eta)=\frac{\psi(x,y)}{\sqrt{u_\infty \nu x}}\, ,
\end{equation}
from which we can express the two dimensionless velocity components:
\begin{equation}
\frac{u}{u_{\infty}}= f'; \qquad \frac{v}{u_\infty} =\frac{1}{2}\sqrt{\frac{u_\infty \nu}{x}} (\eta f' - f)  \, ,
\end{equation}
where $f'$ denotes the first derivative with respect to $\eta$ (same notation is used for higher order derivatives).
Upon substitution of these variables in the continuity and Navier-Stokes equations, we obtain the governing equation for the dimensionless stream function $f(\eta)$:
\begin{equation}
f''' + \frac{1}{2} f f'' = 0\, ,
\label{ge1}
\end{equation}
together with the boundary conditions:
\begin{equation}
 f'(\eta=0)=k f''(\eta=0 )\, ,
\end{equation}
\begin{equation}
f(\eta=0)= 0\, ,
\end{equation}
\begin{equation}
 f'(\eta \rightarrow +\infty)=0\, .
\end{equation}
Considering now the energy equation for the dimensionless temperature $\theta$
\begin{equation}
\theta = \frac{T - T_\infty}{T_w - T_\infty}\, ,
\end{equation}
and using the similarity transformation, the governing equation for the dimensionless temperature becomes:
\begin{equation}
\theta'' + \frac{1}{2} Pr f \theta' = 0\, ,
\label{ge2}
\end{equation}
where $Pr=\nu/\alpha$ is the Prandtl number, and the following boundary conditions are applied:
\begin{equation}
\theta(\eta=0)= 1 \, ,
\end{equation}
\begin{equation}
\theta(\eta \rightarrow +\infty) =0\, .
\end{equation}
The governing equations~\ref{ge1} and \ref{ge2}, which constitute a boundary value problem, are solved numerically via a shooting method which, avoiding the imposition of the boundary condition~\ref{bc3},  stabilizes the computation over a wider range  of $\eta$.
%
%
The equations are solved for different values of $k$, from $k=0$ (no-slip) up to $k=5$, at which the velocity at the wall ($\eta=0$) is $\simeq 70\%$ of the free stream velocity.
The resulting velocity profiles, (rotated by $90$ degrees to be consistent with the sketch of figure \ref{bl}) are shown in figure~\ref{esempi} for different values of $k$ .
Panel~$a$ shows the effect of $k$ on the streamwise component of the velocity,  while panel~$b$ shows the effect of $k$ on the temperature profile.
All the results refer to $Pr=1$, for which the temperature solution can be obtained as $\theta=1-f'$.
For the no-slip case ($k=0$), the Blasius solution (velocity and temperature, shown by the red circles) is recovered.
As expected, by increasing $k$, the amount of slip at the plat increases.
As a consequence, the temperature profiles are also modified, generating larger temperature gradients at the plate.
This corresponds to an heat transfer increase, as also observed in previous studies \citep{martin2006momentum,aziz2014steady}.

Of specific importance in the context of the model developed in the present paper, is the evaluation, as a function of the slip parameter $k$ and for different values of $Pr$, of the ratio between the velocity and the thermal  boundary layer thickness, respectively defined  
\citep{martin2006momentum}:
\begin{equation}
\delta=\int_0^{+\infty} (1-f') d \eta\, , \qquad\text{and} \qquad 
\delta_t=\int_0^{+ \infty} \theta d \eta\, .
\end{equation}

%
%
%
%

\begin{figure}
\begin{picture}(300,170)
\put(45,-20){\includegraphics[width=0.8\columnwidth]{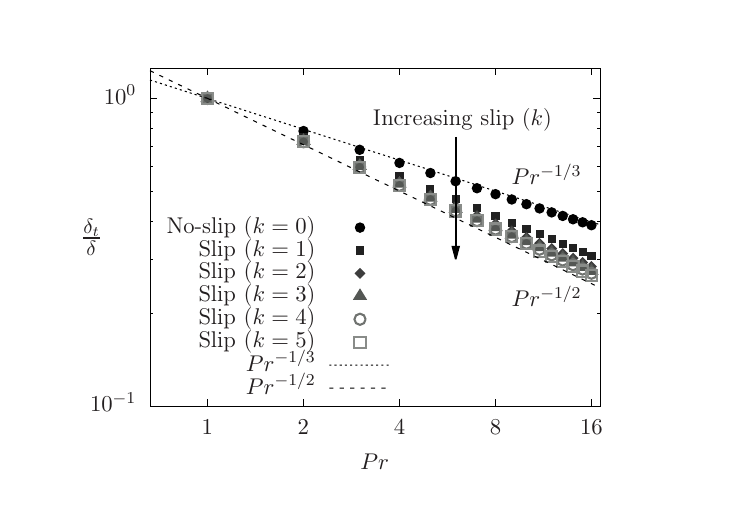}}
\end{picture}
\caption{Ratio between the thermal and momentum boundary layer thickness as a function of the Prandtl number and the slip parameter $k$.
The scaling laws $Pr^{-1/3}$ and $Pr^{-1/2}$ are reported as reference.
Moving from $k=0$ (no-slip) to $k=5$ (slip), for a given value of the Prandtl number, the thermal boundary layer becomes thinner thus leading to an increase of the heat transferred from the wall.}
\label{scalingbl}
\end{figure}

The ratio $\delta_t/\delta$ is shown in figure~\ref{scalingbl} as a function of $Pr$ and for different values of the slip parameter $k$ (different symbols).
We notice that, when the no-slip condition is enforced ($k=0$), the ratio $\delta_t/\delta \sim Pr^{-1/3}$, in agreement with the thermal boundary layer theory on flat plates  \citep{schlichting2016boundary}.
%
%
However, when a slip condition  is introduced at the wall ($k>0$), the ratio $\delta_t/\delta$ relaxes onto the scaling  $\delta_t/\delta \sim Pr^{-1/2}$.
This indicates that, at a given $Pr$, the thermal boundary layer for the slip case becomes thinner compared to the no-slip case, and the heat transfer increases. 
In other words, heat  transfer coefficients  for drops/bubbles (slip surfaces) can be higher compared to the corresponding values for solid particles (no-slip surfaces) \citep{herlina2016isotropic}.
In particular, based on the previous observations, and on the model developed in Sec. \ref{model_sec}, we  can obtain the following  scalings for the heat transfer coefficients:

\begin{equation}
\mathcal{H}^*   \propto Pr^{-2/3}\, \qquad   \text{for no-slip}, 
\end{equation}
\begin{equation}
\mathcal{H}^*   \propto  Pr^{-1/2}\, \qquad   \text{for free-slip}.
\end{equation}

\bibliographystyle{jfm}
\bibliography{jfm}

\end{document}